\documentclass[reprint,aps,showpacs,prx,superscriptaddress,floatfix,10pt,nofootinbib]{revtex4-2}
\usepackage[T1]{fontenc}
\usepackage[utf8]{inputenc}
\usepackage{amsmath}
\usepackage{amssymb}
\usepackage{esint}
\usepackage{graphicx}
\usepackage{bbold}
\usepackage{appendix}
\usepackage{bibunits}
\usepackage{verbatim}
\usepackage{bm}
\usepackage{xr-hyper}
\usepackage{lipsum} 

\makeatletter

\usepackage{color}

\usepackage{physics} 
\usepackage{xcolor} 

\definecolor{mygreineren}{RGB}{0, 126, 0}
\definecolor{myorange}{RGB}{255, 136, 19}
\definecolor{mymagenta}{RGB}{200, 0, 100}

\def\@fnsymbol#1{\ensuremath{\ifcase#1\or \dagger\or \ddagger\or
   \mathsection\or \mathparagraph\or \|\or **\or \dagger\dagger
   \or \ddagger\ddagger \else\@ctrerr\fi}}
    \makeatother

\makeatother

\begin{document}
\title{The charge-singlet measurement toolbox}


\author{Abhijit Chakraborty}
\affiliation{Institute for Quantum Computing, University of Waterloo, Waterloo, ON, Canada, N2L 3G1}
\affiliation{Department of Physics \& Astronomy, University of Waterloo, Waterloo, ON, Canada, N2L 3G1}





\author{Randy Lewis}
 \affiliation{Department of Physics and Astronomy, York University, Toronto, ON, Canada, M3J 1P3}


\author{Christine A. Muschik}
\affiliation{Institute for Quantum Computing, University of Waterloo, Waterloo, ON, Canada, N2L 3G1}
\affiliation{Department of Physics \& Astronomy, University of Waterloo, Waterloo, ON, Canada, N2L 3G1}
\affiliation{Perimeter Institute for Theoretical Physics, Waterloo, ON, Canada, N2L 2Y5}


\date{\today}



\begin{abstract}
Symmetry is fundamental to physical laws across different scales—from spacetime structure in general relativity to particle interactions in quantum field theory. Local symmetries, described by gauge theories, are central to phenomena such as superconductivity, topological phases, and the Standard Model of particle physics. Emerging simulation techniques using tensor network states or quantum computers offer exciting new possibilities of exploring the physics of these gauge theories, but require careful implementation of gauge symmetry and charge-neutrality constraints. This is especially challenging for non-Abelian gauge theories such as quantum chromodynamics (QCD), which governs the strong interaction between quarks and gluons. In a recent article (arXiv:2501.00579), we introduced ``charge-singlet measurements'' for quantum simulations, consisting of a projection based technique from group representation theory that allowed us to probe for
the first time the phase diagram of (1+1)-dimensional QCD on a quantum computer. In this article, we show more broadly how to apply charge-singlet measurements as a flexible tool for both classical and quantum simulations of discrete and continuous gauge theories. Our approach extends the use of charge-singlet measurements beyond state preparation in the charge neutral (charge-singlet) sector to include noise mitigation in symmetry-preserving time-evolution circuits. We further demonstrate how this method enables the computation of thermodynamic observables--such as entropy--within the charge-singlet subspace, providing a new tool for exploring the connection between quantum thermodynamics and gauge symmetry.
\end{abstract}
\maketitle

\section{Introduction} 
In most branches of science, understanding the symmetries of a system is essential. Symmetries reveal invariances under specific transformations, which, by Noether’s theorem, correspond to conserved quantities. These conservation laws impose constraints on the equations of motion, enabling efficient and accurate descriptions of the system \cite{schwichtenberg2018physics}. Just as the presence of symmetry provides critical insight, the breaking of symmetry also carries equally important information—frequently signaling phase transitions or emergent behavior \cite{beekman2019introduction}. Phenomena such as superconductivity, ferromagnetism, and topological phases in condensed matter physics, as well as hadron mass generation and the Higgs mechanism in particle physics, are deeply rooted in the breaking of fundamental symmetries. For this reason, identifying and implementing a system’s symmetries is a vital step in any attempt to model or simulate its properties.

Local symmetries, associated with discrete or continuous groups, play a fundamental role in both condensed matter physics and particle physics. Especially, the Standard Model of particle physics is described by the local gauge symmetry SU(3)$\times$ SU(2)$\times$ U(1), each group describing a particular sector of the model \cite{georgi2000lie}. The strong interaction between fundamental fermionic particles (quarks) mediated by gauge fields (gluons) follow the symmetry group SU(3), i.e., the system remains invariant under local gauge symmetry transformations which are elements of the non-Abelian SU(3) group. The theory of quarks and gluons, quantum chromodynamics (QCD), has the potential to give us significant insights about the phases of matter at the early stage of the universe and within compact stars. Calculating scattering cross sections in QCD is especially hard because perturbative methods fail due to strong coupling strength. 
A non-perturbative approach to solving QCD is to discretize spacetime into a lattice, simulate the system numerically, and recover continuum physics through extrapolation—a method known as lattice QCD \cite{gupta_introduction_1998,philipsen2013qcd,rothe2012lattice}. 

The traditional approach to simulating lattice QCD is a Lagrangian-based formalism, which, despite its phenomenal success, encounters significant difficulty at finite matter densities due to sign problems \cite{nagata2022finite,muroya2003lattice}. While different methods have been explored to overcome these sign problems, avoiding them efficiently for a wide range of matter densities still remains an open challenge \cite{guenther2021overview,alexandru2022complex,nagata2022finite,ratti2018lattice,achenbach2024present}. This leaves a large region in the temperature-chemical potential phase diagram of QCD out of our access. Furthermore, real-time dynamics simulation of lattice QCD is also hard in the Lagrangian-based approach due to the euclideanization of the time coordinate. 
More recently, tensor networks and quantum computers have used the Hamiltonian approach \cite{Wilson1974sk,gregory2000hamiltonian} to simulate lattice gauge theories (LGT), 	 	
circumventing the barrier of the sign problem \cite{magnifico2024tensor,banuls2023tensor,di2024quantum,bauer2023quantum,bass2022quantum}. Advances in quantum hardware have enabled the implementation of proof-of-concept protocols for simulating dynamics and preparing eigenstates of lattice gauge theories on existing quantum devices \cite{martinez_real-time_2016, rahman2022, martinez_compiling_2016,atas2021,paulson2021simulating,atas2021,atas2023,yang_observation_2020,farrell2024scalable,kavaki2024square,nguyen_digital_2022,farrell2024quantum,davoudi2024scattering,klco_su2_2020,ciavarella_trailhead_2021,zhou2022thermalization,ciavarella2024string,su2023observation,de2010simulating,mueller2023quantum,turro2024qutrit,mildenberger2025confinement,halimeh2025cold,schuhmacher2025observation,kavaki2025false,farrell2025digital,davoudi2025quantum}. Although most of these simulations are for the much simpler but physically rich and relevant toy models in (1+1)-D, first steps towards higher dimensional LGTs have also been taken recently \cite{mueller2025quantum,meth2025simulating,osborne2025large,cochran2024visualizing,gonzalez2025observation,crippa2024analysis,gyawali2024observation}.

A significant challenge for the Hamiltonian-based approaches like tensor networks and quantum computation is the implementation of the gauge symmetries \cite{zohar_formulation_2015}. In particular, in the absence of background fields, all physical states in a lattice gauge theory must satisfy not only the local Gauss's law constraints imposed by the gauge symmetry but also a global charge neutrality condition. For an SU($N_c$) system, this requires that the total charge corresponding to each conserved generator—commonly referred to as a color charge, must vanish. Such a state of the system is called a charge-singlet state. This leads to a set of $N_c^2-1$ operator-valued constraints that need to be implemented in any simulation of a SU($N_c$) LGT. In the literature, several methods have been proposed to implement such constraints, including penalty-based techniques \cite{halimeh2022stabilizing,gaz2025quantum,davoudi2023towards} and symmetry-preserving circuit \cite{atas2021,kokail_self-verifying_2019} or network architectures \cite{luo2021gauge}—each with its own set of advantages and limitations. In \cite{than2024phase}, we introduced an alternative strategy based on projection techniques from group representation theory to explore the phase diagram of (1+1)-dimensional QCD using a trapped-ion quantum computer. Unlike other approaches that involve preparing a charge-singlet state on the quantum device, our method employs a measurement protocol designed to yield the correct expectation values of observables as if measurements were performed on a charge-singlet state. Thus, the color-neutrality constraint is enforced implicitly through a postprocessing step applied at the readout stage, following the completion of state preparation. We note that preparing a charge-singlet state on the device is not required at any stage of the protocol to obtain the correct expectation value within the singlet subspace. Instead, the desired result is achieved through modified measurements applied to the non-charge-singlet state generated on the device. One significant advantage of using this projection method or charge-singlet measurements (CSM), is that we are free to choose any architecture for our network or circuit for the state preparation part without considering consequences of not satisfying the gauge symmetry constraints.

In this article, we present a broader perspective on how CSMs can serve as a versatile toolbox for quantum simulations and tensor network computations. We begin by reviewing the color-neutrality constraints that define charge-singlet states in Sec.~\ref{sec:problem-statement}. Next, we outline the theoretical framework underlying the CSMs employed in \cite{than2024phase}, and generalize it to a wider class of lattice gauge theories (LGTs), including those with discrete gauge symmetries (Sec.~\ref{sec:theory-framework}). We then derive the expression for the diagonal projection operator used in the CSM method and explicitly demonstrate the connection between conserved charges and the irreducible representations of the underlying gauge group, thereby providing further insight into the projection-based approach (Sec.~\ref{sec:diagonal-projector}).
In \cite{than2024phase}, this technique was applied to measure the chiral condensate in (1+1)-dimensional SU(2) and SU(3) LGTs at finite temperature and chemical potential. Building on this, we show how CSMs can be used to extract the expectation value of the electric field Hamiltonian (dependent on color electric field) at finite temperature (Sec.\ref{sec:thermal-states}).
We further explore other application areas of CSMs, by proposing their use in calculating thermodynamic quantities in LGTs (Sec.~\ref{sec:GI-entropy}) and by demonstrating their potential for mitigating noise effects in quantum simulations of LGTs on near-term quantum hardware (Sec.\ref{sec:noise-mitigation}). Additionally, as an extension of the discussion in Sec.\ref{sec:diagonal-projector}, we demonstrate how the projection operator can be used to determine the dimension of the charge-singlet subspace (Sec.\ref{sec:dimension_subspace}).
We conclude with a discussion on the broader applicability of this CSM-based toolbox for future simulations of lattice gauge theories.

\section{Problem statement and overview}\label{sec:problem-statement}
In this article, we study non-Abelian lattice gauge theories that are relevant to the Standard Model of particle physics. While Abelian gauge theories—commonly encountered in both condensed matter and particle physics—lead to algebraic constraints that are generally simpler to implement, non-Abelian theories are inherently more complex. This complexity arises because their conserved charges are represented by non-commuting operators, resulting in operator-valued constraints. Although the theoretical framework we present is broadly applicable (including to Abelian cases), our focus is specifically on SU(2) and SU(3) lattice gauge theories with fermionic matter \cite{zohar_formulation_2015}, motivated by their direct connection to QCD. The Hamiltonian in this case consists of the following contributions
\begin{equation}
    \hat{H} = \hat{H}_{kin} + m  \hat{H}_{m} + g^2\hat{H}_{el} + \frac{1}{g^2} \hat{H}_{mag} - \mu \hat{H}_{chem} \;,\label{eq:generic-Hamiltonian}
\end{equation}
where $ \hat{H}_m$ is the mass term, $ \hat{H}_{el}$ is the electric field term, $ \hat{H}_{mag}$ is the magnetic field term, $\hat{H}_{kin}$ is the kinetic energy (pair-creation) operator, $H_{chem}$ is the chemical potential term, and $m, \mu, g$ are the dimensionless mass parameter, chemical potential, and coupling strength. In (1+1)-D, the magnetic term $\hat{H}_{mag}$ is absent. The explicit forms of the terms appearing in Eq.~(\ref{eq:generic-Hamiltonian}) for (1+1)-D SU(2) and SU(3) LGTs are given in \cite{atas2021,atas2023} and in Appendix \ref{app:unit-cell-hamiltonian} for convenience.

The Hamiltonian $\hat{H}$ commutes with all the generators of the gauge group. The generators are thus conserved quantities, and are called the conserved color charges of the system. For a SU($N_c$) LGT, there are $N_c^2-1$ conserved charges. Local gauge symmetry implies that at each vertex of the lattice the color charges satisfy the Gauss's laws. In (1+1)-D with open boundary conditions, these local Gauss's laws can be integrated out as the gauge field configurations can be determined from the color charge distribution on the vertices \cite{atas2021,atas2023}. Apart from the local Gauss's laws, in the absence of background charges, a physical state should have zero net charge for all $N_c^2-1$ color charges. This choice of the absence of background charges is inspired from the physical observation that we can only observe color-neutral composite particles in the Standard Model. The state $\ket{\psi_0}$, which we call a color-singlet or charge-singlet state, is then defined by
\begin{equation}
    \hat{Q}_{tot}^j \ket{\psi_0} = 0 \;,\qquad\forall j \in \{1,2,\dots,N_c^2-1\}\;,
    \label{eq:color-neutral-condition}
\end{equation}
where $\hat{Q}^j_{tot}$ is a total color charge operator. Not all states $\ket{\psi}$, which can be defined on the Hilbert space, will satisfy this condition. To give a concrete example, we consider the SU(3) unit cell in (1+1)-D in the staggered Kogut-Susskind formalism \cite{kogut_hamiltonian_1975}. We can define a state $\ket{\bar{r}vvrvv}$ on a unit cell, which is just the presence of a red particle $r$ (quark) and a red antiparticle $\bar{r}$ (antiquark) with the other sites being vacant ($v$). This state is a valid quantum state belonging to the Hilbert space of a unit cell but does not satisfy the color-singlet condition (\ref{eq:color-neutral-condition}) (see Appendix~\ref{appsec:singlet-states} for examples of singlet states defined on a unit cell).  
  
In order to simulate the physics of a LGT in the color-neutral sector on a quantum device, we need to ensure that observables are measured on a state that satisfies the charge-singlet condition along with the local Gauss's laws. However, satisfying this condition is often hard, especially in a variational protocol. In a variational state preparation protocol (ground, excited, or thermal state), enforcing the constraints usually involves using a symmetry-preserving ansatz or adding a penalty term to the cost function for each conserved charge.

Both approaches are quite non-trivial and especially demanding for thermal states, for which the initial states are not necessarily color-neutral either. Moreover, in the case of symmetry preserving circuits, or time evolution (where color-neutrality is preserved due to the time-evolution operator $e^{-i\hat{H}t}$ commuting with the global color charges), noisy gates often lead to a final state which does not belong to the charge-singlet subspace anymore.

In this article, we demonstrate the use of a group-theoretical projection technique to address this issue of ensuring the set of constraints in Eq.~(\ref{eq:color-neutral-condition}) along with the local Gauss's law constraints. We start by noticing the fact that when mapped to spins, the Hamiltonian $\hat{H}$ in Eq.~(\ref{eq:generic-Hamiltonian}) is a matrix which belongs to a reducible representation of dimension $2^{N_cN}\times2^{N_cN}$ of the group SU($N_c$), where $N$ is the number of spatial sites of the lattice. All charge-singlet states satisfying the condition Eq.~(\ref{eq:color-neutral-condition}) then belong to the singlet representation by definition. To obtain the color-neutral state from a state defined on the reducible representation we need to project it to the singlet representation, which is a well-defined technique in group representation theory. 

The theoretical framework of this technique is explained in the next section with the focus on SU(2) and SU(3) LGT in (1+1)-D with open boundary conditions. For this setting, we explicitly calculate the projection operator and explain how it can be used to get expectation values of observables on the singlet subspace without specifically preparing the singlet state. A diagonalized form of the projection operator is then used to find the expectation value of an observable by performing a set of modified measurements on the reducible state prepared on the device. Delegating the implementation of the color-neutrality condition to the readout stage inspired our nomenclature of this method `charge-singlet measurements'. We thus trade preparing singlet states on the quantum device for the ability to compute correct expectation values of observables through post-processing.

We also emphasize the connection of the projection method with thermodynamic quantities, such as entropy, in the context of LGTs (Sec.~\ref{sec:GI-entropy}). In the presence of noise, this method can be used to project out certain non-charge-singlet noise components in symmetry-preserving and time-evolution circuits, leading to more accurate estimates of observables (Sec.~\ref{sec:noise-mitigation}). The projection method thus acts as an effective noise mitigation strategy, which can be used in conjunction with other ways of mitigating noise \cite{cai2023quantum}. Not only is this method relevant for quantum computers, it is also very useful for classical Hamiltonian-based simulation of LGTs too. For instance, instead of explicitly constructing tensor network architectures satisfying the gauge constraints, one can employ the projection operator to efficiently compute relevant observables. We present use cases for each of the aforementioned applications to illustrate the advantages of this method.

\section{Theoretical framework for charge-singlet measurements}\label{sec:theory-framework}
The underlying framework for charge-singlet measurements follows from the representation theory of groups. In our specific case of SU(2) and SU(3) LGT in (1+1)-D, the underlying gauge group is a non-Abelian Lie group. However, the projector formalism that we used in \cite{than2024phase} can be applied to more general symmetry groups, including discrete groups as well. In this section, we briefly explain the theoretical background for the determination of the projection operator, and its application to SU(2) and SU(3) LGT. In the following, we initially use notation standard in group theory and subsequently adopt the notation commonly used in quantum mechanics.

\subsection{General form of the projector}\label{sec:general-projector}
For a finite group $G$, we consider a reducible representation $D: G\rightarrow GL(V)$, where $V$ is the underlying vector space, and $GL(V)$ is the general linear group of the vector space $V$. The vector space $V$ can then be decomposed as $V = \bigoplus_i n_iV_i$, where a particular representation $V_i$ appears $n_i$ times. Given a vector $v\in V$, we can then define the projection operator \cite{cornwell1997group} 
\begin{equation}
    P_i = \frac{{\rm dim}(V_i)}{|G|} \sum_{g\in G} \;\chi^i(g)^* D(g) \;,
\end{equation}
such that $P_iv$ belongs to the subspace $V_i$. Here, $|G|$ is the order of the group, and $\chi^i(g)$ is the group character corresponding to the representation $V_i$, defined by
\begin{equation}
    \chi^i(g) := {\rm Tr}(D^i(g)) \;.
\end{equation}
For a singlet representation, the character is trivial, i.e. $\chi^0(g) = 1$. So, the projection operator for the singlet representation is defined as
\begin{equation}
    P_0 = \frac{1}{|G|} \sum_{g\in G} \; D(g) \;. \label{eq:P_0-general-finite}
\end{equation}
The projection operator defined in this way has only 0 and 1 as eigenvalues (as customary for an idempotent operator). Thus the trace of the operator $P_0$ yields the number of copies $n_0$ of the singlet decomposition that appear in $V$.

Eq.~(\ref{eq:P_0-general-finite}) can be generalized to continuous groups as well. We are primarily interested in compact non-Abelian Lie groups SU(2) and SU(3), which appear naturally in the Standard Model of particle physics. The underlying vector space $V$ in our case is the Hilbert space where the quantum state vector $\ket{\psi}$ is defined and the Hamiltonian in Eq.~(\ref{eq:generic-Hamiltonian}) acts on. For non-Abelian lattice gauge theories in the absence of background charges, the physical state must be color-neutral and belong to the singlet representation. However, a general state $\ket{\psi}$ can belong to a reducible representation of the underlying gauge group. Given a reducible representation, the projection operator $\hat{P}_0$ can be used to extract the singlet state. From here, we denote operators with a hat, to be consistent with the notations used in quantum mechanics.

For a continuous gauge group, Eq.~(\ref{eq:P_0-general-finite}) takes the form
\begin{equation}
    \hat{P}_0 = \int d\mu(\boldsymbol{\alpha})\; \hat{U}(\boldsymbol{\alpha}) \;, \label{eq:P_0-general-Lie}
\end{equation}
where $\hat{U}(\boldsymbol{\alpha})$ is an element of the group belonging to the reducible representation and parametrized by a set of parameters $\boldsymbol{\alpha}$. $d\mu(\boldsymbol{\alpha})$ is the normalized group integral measure dependent on the parameters $\boldsymbol{\alpha}$, i.e., $\int d\mu(\boldsymbol{\alpha}) = 1$. This is the continuum equivalent of the normalized sum $|G|^{-1}\sum_{g\in G}$ in Eq.~(\ref{eq:P_0-general-finite}).

We note here that $\hat{P}_0$ is a proper projection operator, i.e., $\hat{P}_0^2 = \hat{P}_0$ and all of its eigenvalues are either 0 or 1. So, for a given quantum state $\ket{\psi}$ belonging to a reducible representation, its projection onto the singlet subspace is given by 
\begin{equation}
    \ket{\psi_0} = \frac{\hat{P}_0\ket{\psi}}{\bra{\psi}\hat{P}_0 \ket{\psi}} \;.
\end{equation}
The denominator ensures that the projected quantum state vector $\ket{\psi}_0$ is normalized to unity.
For a density matrix $\hat{\rho}$, the corresponding projection to the singlet subspace is given by
\begin{equation}
    \hat{\rho}_0 = \frac{\hat{P}_0 \hat{\rho}\hat{P}_0}{{\rm Tr}(\hat{\rho}\hat{P}_0)} \;.\label{eq:singlet-density-projection-od}
\end{equation}
Once again, the denominator here enforces the unit-trace condition of the projected density matrix $\hat{\rho}_0$.

Given a state $\hat{\rho}$ defined on the full Hilbert space (reducible representation), we can now use Eq.~(\ref{eq:singlet-density-projection-od}) to calculate the expectation value of any observable $\hat{O}$
\begin{equation}
    {\rm Tr}(\hat{\rho}_0\hat{O} ) = \frac{{\rm Tr}(\hat{\rho}\hat{O}\hat{P}_0)}{{\rm Tr}(\hat{\rho}\hat{P}_0)}\;, \label{eq:exp-val-P_0}
\end{equation}
where we have used the idempotent property of the projection operator $\hat{P}_0^2 = \hat{P}_0$ and the commutation relation $[\hat{O},\hat{P}_0] = 0$. This follows from the fact that all observables commute with the gauge group generators and $\hat{P}_0$ is defined in terms of the generators as shown in Eq.~(\ref{eq:P_0-general-Lie}). For any experimental study of a lattice gauge theory, our ultimate goal is to measure some observable on the singlet subspace, which is exactly the left-hand side of Eq.~(\ref{eq:exp-val-P_0}). If we do not enforce the charge-singlet condition at the state preparation stage, Eq.~(\ref{eq:exp-val-P_0}) can be used for extracting the desired expectation value in the charge-singlet subspace.

\subsection{Reduction to diagonal form}
While Eq.~(\ref{eq:exp-val-P_0}) is valid for calculating the expectation value of an observable, it is hard to implement on a quantum device. The projection operator $\hat{P}_0$ is non-diagonal due to the presence of arbitrary group elements $\hat{U}(\boldsymbol{\alpha})$ in the definition of $\hat{P}_0$. The parametrization of $\hat{U}(\boldsymbol{\alpha})$ in terms of the group generators will include both diagonal and non-diagonal generators. As a result, the Pauli-string decomposition of the operator $\hat{P}_0$ will have many non-diagonal Pauli operators. To avoid these non-diagonal measurements, we note that every element of a Lie gauge group can be diagonalized using a similarity transformation $\hat{U} = \hat{U}_s \hat{U}_d \hat{U}_s^\dagger$, where $\hat{U}_s$ is the similarity transformation and $\hat{U}_d$ is the diagonalized form of the operator. Both $\hat{U}_d$ and $\hat{U}_s$ generally depend on the parameter values, so each $\boldsymbol{\alpha}$ specifies a distinct $\hat{U}_d$ and $\hat{U}_s$. Using Eq.~(\ref{eq:P_0-general-Lie}) and Eq.~(\ref{eq:exp-val-P_0}), we then obtain
\begin{align}
    {\rm Tr}(\hat{\rho}\hat{O}\hat{P}_0) &= \int_{\boldsymbol{\alpha}\in SU(N_c)} d\mu(\boldsymbol{\alpha})\;{\rm Tr}(\hat{\rho}\hat{O}\hat{U}_s(\boldsymbol{\alpha})\hat{U}_d(\boldsymbol{\alpha})\hat{U}_s^\dagger(\boldsymbol{\alpha})) \nonumber\\
    &= \int_{\boldsymbol{\alpha}\in SU(N_c)} d\mu(\boldsymbol{\alpha})\;{\rm Tr}(\hat{\rho}\hat{O}\hat{U}_d(\boldsymbol{\alpha})) \\
    & = {\rm Tr}(\hat{\rho}\hat{O}\hat{K})\;,
\end{align}
where we have used the fact that both $\hat{\rho}$ and $\hat{O}$ commute with any group element $\hat{U}(\boldsymbol{\alpha})$. We have also defined  
\begin{equation}
    \hat{K} =\int_{\boldsymbol{\alpha}\in SU(N_c)} d\mu(\boldsymbol{\alpha})\; \hat{U}_d(\boldsymbol{\alpha}) \;.\label{eq:K-first-def}
\end{equation}
Since $\hat{U}_d(\boldsymbol{\alpha})$ is completely diagonal, we can use a parametrization that only covers the diagonal subgroup, i.e., the Cartan subgroup $\mathcal{C}_{N_c}$ of the SU($N_c$) group. The diagonal group element can then be written using only the diagonal generators of the group, viz., the diagonal conserved charges \cite{leyaouanc1989,elze1986quantum,mclerran1985thermodynamics}. In the case of (1+1)-D LGT with open boundary conditions, the diagonal group elements can be expressed in terms of the total diagonal charges.
\begin{align}
    \hat{K} = \int_{\alpha_j,\hat{Q}^j_{tot}\in\mathcal{C}_{N_c}} d\mu(\boldsymbol{\alpha}) \,e^{i\sum_j\alpha_j \hat{Q}_{tot}^{\rm j}} \;.
    \label{eq:K_diagonal_general}
\end{align}
where $\hat{Q}_{tot}^j$ are total diagonal generators belonging to the Cartan subalgebra of the $su(N_c)$ algebra.

We note here that the operator $\hat{K}$ is completely diagonal as it is defined in terms of the diagonal group elements only. So, any Pauli decomposition of $\hat{K}$ will only contain diagonal Pauli strings. However, a consequence of the diagonalization is that $\hat{K}$ is not a projection operator in the conventional sense, i.e., it is not idempotent. Due to the integral in Eq.~(\ref{eq:K-first-def}) and a parametrization defined specifically for elements in the Cartan subalgebra, we lose the idempotent property that was true for $\hat{P}_0$. Consequently, in Eq.~(\ref{eq:singlet-density-projection-od}), we cannot replace $\hat{P}_0$ with $\hat{K}$ to recover the charge-singlet density matrix. We can only replace $\hat{P}_0$ with $\hat{K}$ under a trace sign with an observable. This means, we can use the operator $\hat{K}$ to recover the expectation value of any observable:
\begin{equation}
    \langle \hat{O}\rangle_{0} =  {\rm Tr}(\hat{\rho}_0\hat{O} ) = \frac{{\rm Tr}(\hat{\rho}\hat{O}\hat{K})}{{\rm Tr}(\hat{\rho}\hat{K})}\;, \label{eq:exp-val-K}
\end{equation}
but we do not gain access to the charge-singlet state. The advantage is that if the observable $\hat{O}$ is diagonal as well, then we can evaluate Eq.~(\ref{eq:exp-val-K}) using only computational basis measurements. For completeness, we provide an alternate derivation of Eq.~(\ref{eq:exp-val-K}) in Appendix \ref{app:alternative-deriv} following \cite{greiner2012quantum,elze1986quantum}), which motivates the use of the diagonal form $\hat{K}$ from another perspective.

\subsection{Projection operator for a general discrete non-Abelian group}\label{sec:discrete-group}
The derivation and use of the diagonal form of the projection operator $\hat{K}$ in Eq.~(\ref{eq:K_diagonal_general}), relies on the fact that in (1+1)-D LGTs with open boundary conditions, $\hat{U}(\boldsymbol{\alpha})$ can be expressed in terms of the total charge generators $\hat{Q}_{tot}^j$, as the local Gauss's laws are automatically satisfied due the gauge field elimination. However, for (1+1)-D LGT with periodic boundary conditions and generally for spatial dimensions higher than 1D, not all gauge fields can be eliminated, and the residual gauge degrees of freedom must be explicitly accounted for. To enable quantum simulation, the remaining infinite-dimensional gauge fields are usually truncated and mapped to spin degrees of freedom suitable for current quantum hardware. This truncation gives rise to a finite-dimensional gauge group. Motivated by this requirement, we provide here a general description for constructing the projection operator for a finite non-Abelian gauge group. In this case, Eq.~(\ref{eq:P_0-general-finite}) and Eq.~(\ref{eq:singlet-density-projection-od}) still remain valid, only the expression of the projection operator is tailored to a non-Abelian finite gauge group. 

We consider a finite non-Abelian gauge group $G$ and a lattice in (2+1)-D or (3+1)-D. The set of vertices $\mathcal{V}$ host fermions and the edges between vertices host the gauge link operator $\hat{U}_{\boldsymbol{n},\boldsymbol{n} + \hat{k}}$ with $\boldsymbol{n}$ representing the indices of a vertex $v \in \mathcal{V}$. Since we are considering a local gauge symmetry, we can define a gauge transformation $\hat{\Theta}_g^v$ on each vertex $v$ of the lattice corresponding to a group element $g \in G$ as follows \cite{zohar_formulation_2015}
\begin{equation}
    \hat{\Theta}_g^v = \left(\prod_{o,i}\hat{\Theta}_{g,o}^{L,v}\,\hat{\Theta}_{g,i}^{R,v} \right)\hat{\Theta}_{g}^{Q,v}\;. \label{eq:general-gauge-transformation}
\end{equation}
We explain the notation introduced in the equation above. $\hat{\Theta}_{g,o}^{L,v}$ and $\hat{\Theta}_{g,i}^{R,v}$ are gauge transformations that act on the link operators $\hat{U}$ which reside on the outgoing (denoted by subscript $o$) and ingoing link (denoted by subscript $i$) of vertex $v$, respectively. Their action on a link operator $\hat{U}$ is given by
\begin{align}
    (\hat{\Theta}_{g}^{L})^j \,\hat{U}_{mn}^j\, (\hat{\Theta}_{g}^{L\dagger})^j &= D^{j}_{mk} (g^{-1}) \hat{U}_{kn}^j \;,\\
    (\hat{\Theta}_{g}^{R})^j \,\hat{U}_{mn}^j\, (\hat{\Theta}_{g}^{R\dagger})^j &= \hat{U}_{mk}^j D^{j}_{kn} (g) \;,
\end{align}
where $D^{j}(g)$ is the $j$-representation of the group $G$, the same representation that is used to write the link operators $\hat{U}$ and the gauge transformation operators. $(\hat{\Theta}_{g}^{L})^j$ and $(\hat{\Theta}_{g}^{R})^j$ corresponds to the left action and right action representation of the non-Abelian group, which are related via the adjoint representation and can be written explicitly in terms of group representation elements $D^j(g)$ \cite{zohar_formulation_2015,atas2021}. 
$(\hat{\Theta}_{g}^{Q})^j$ is the gauge transformation for a fermionic operator on the vertex, which transforms the fermionic operators in the following way
\begin{align}
    (\hat{\Theta}_{g}^{Q})^j \,\hat{\phi}_{a}\, (\hat{\Theta}_{g}^{Q\dagger})^j &= D^{j}_{ab} (g^{-1})\, \hat{\phi}_{b} \;,\\
    (\hat{\Theta}_{g}^{Q})^j \,\hat{\phi}_{a}^\dagger\, (\hat{\Theta}_{g}^{Q\dagger})^j &=  \hat{\phi}_{b}^\dagger\, D^{j}_{ba} (g) \;,
\end{align}
where $\hat{\phi}_{a}$ is the fermionic operator. The transformation operator for the charges $(\hat{\Theta}_{g}^{Q})^j$ is given by \cite{zohar_formulation_2015}
\begin{equation}
    (\hat{\Theta}_{g}^{Q})^j = e^{i\hat{\phi}_{a}^\dagger q_{ab}^j (g)\hat{\phi}_b}\,{\rm det}(D^j(g^{-1}))^k \,,
\end{equation}
where $q^j(g) = -i\log(D^j(g))$ and $k=1$ (0) for the vertex $v$ in an odd (even) sublattice.

We have the definition of all the terms in Eq.~(\ref{eq:general-gauge-transformation}). A singlet state is then defined as a state $\ket{\psi_{0}}$ which remains invariant under the transformation Eq.~(\ref{eq:general-gauge-transformation}) for all the vertices on the lattice, i.e.,
\begin{equation}
    \hat{\Theta}_{g}^{v} \ket{\psi_{0}} = \ket{\psi_{0}}\;, \label{eq:phys-state-def}
\end{equation}
where we have suppressed the index $j$ denoting the representation.

We note that Eq.~(\ref{eq:general-gauge-transformation}) is true for all gauge groups. For continuous groups, the gauge transformation operators can be written in terms of generators of the gauge group,
\begin{equation}
    \hat{\Theta}_{g}^{v} = e^{i\boldsymbol{\alpha}_g\cdot \hat{\boldsymbol{G}}_v}\;,\qquad \hat{\boldsymbol{G}}_v=\sum_i \hat{\boldsymbol{R}}_{i,v} + \sum_o \hat{\boldsymbol{L}}_{o,v} + \hat{\boldsymbol{Q}}_v\;,
\end{equation}
where the sum over $i$ and $o$ denote sum over the number of incoming and outgoing links and $\hat{\Theta}_{g,o}^{L} = e^{i\boldsymbol{\alpha}_g\cdot\hat{\boldsymbol{L}}_o}$, $\hat{\Theta}_{g,o}^{R} = e^{i\boldsymbol{\alpha}_g\cdot\hat{\boldsymbol{R}}_o}$, and $\hat{\Theta}_{g}^{Q} = e^{i\boldsymbol{\alpha}_g\cdot\hat{\boldsymbol{Q}}}$. The condition for a physical state is then it has to be the eigenstate of the operator $\hat{\boldsymbol{G}}_v$ with zero eigenvalue for all $v$. In (1+1)-D with open boundary conditions, after gauge field eliminations, this condition becomes equivalent to Eq.~(\ref{eq:color-neutral-condition}) where the total color charges $\hat{Q}_{tot}^j$ are completely written in terms of fermionic operators.

However, this is no longer true for a discrete gauge group. Therefore, the Gauss's laws cannot be expressed explicitly in terms of generators only. We can still define the projection operator for the discrete gauge group from Eq.~(\ref{eq:P_0-general-finite}) as
\begin{equation}
    \hat{P}_0 = \prod_{v} \frac{1}{|G|} \sum_g \hat{\Theta}_{g}^v\;,
\end{equation}
where $\hat{\Theta}_{g}^v$ is given by Eq.~(\ref{eq:general-gauge-transformation}).
The explicit form of $\hat{P}_0$ in terms of Pauli operators will then depend on the specific encoding of the gauge transformation onto qubits or qudits \cite{gonzalez-cuadra_hardware_2022,ballini2024symmetry} and on the finite gauge group resulting from the truncation of the infinite-dimensional gauge operators.



\section{Diagonal projection operator $\hat{K}$ for SU(2) and SU(3) LGTs in (1+1)-D}\label{sec:diagonal-projector}
With the general framework described in Sec.~\ref{sec:theory-framework}, we now focus on the non-Abelian gauge groups relevant to the Standard Model of particle physics, viz., SU(2) and SU(3) gauge groups. In this section, we provide an explicit description of how the diagonal projection operator $\hat{K}$ can be determined analytically for (1+1)-D SU(2) and SU(3) LGT with open boundary conditions.

\subsection{Conserved charges}
As expressed in Eq.~(\ref{eq:K_diagonal_general}), the projection operator $\hat{K}$ is defined in terms of the diagonal conserved charges. For SU(2) and SU(3) LGT in (1+1)-D with open boundary conditions, conserved charges can be written in terms of only fermionic operators after eliminating all the gauge fields. For higher dimensions or periodic boundary conditions, the remaining gauge field needs to be truncated and would lead to a discrete group. The general framework for a discrete group is described in Sec.~\ref{sec:discrete-group}.

SU(2) LGT involves only one diagonal conserved charge, which we denote by $\hat{Q}^z_{tot}$. The explicit expression of $\hat{Q}^z_{tot}$ follows from the diagonal generator $\hat{\sigma}^z/2$ of the SU(2) gauge group and after a Jordan Wigner transformation can be written in terms of Pauli operators only \cite{atas2021} 
\begin{align}
    \hat{Q}^z_{tot} = \sum_{n=1}^N  \hat{Q}^z_n 
    &= \frac{1}{2}\sum_{n=1}^N \sum_{ij} \hat{\phi}_n^{i\,\dagger} (\hat{\sigma}^z)_{ij} \,\hat{\phi}_n^{j} \nonumber\\
    &= \frac{1}{4}\sum_{n=1}^N \left(\hat{\sigma}^z_{2n-1} - \hat{\sigma}^z_{2n} \right) \label{eq:Qz_tot}\;.
\end{align}
Here $\hat{\phi}_n^{i}$ are the staggered fermionic annihilation operator, and $N$ is the number of spatial sites of the staggered lattice (for a unit cell $N=2$).

For SU(3), there are two diagonal generators of the group, which give us two diagonal total conserved charges of the system \cite{atas2023}:%
{\allowdisplaybreaks
\begin{align}
    \hat{Q}^3_{tot} = \sum_{n=1}^N  \hat{Q}^3_n &= \frac{1}{2}\sum_{n=1}^N \sum_{ij} \hat{\phi}_n^{i\,\dagger} (\hat{\lambda}^3)_{ij} \,\hat{\phi}_n^{j} \nonumber\\
    &= \frac{1}{4}\sum_{n=1}^N \left(\hat{\sigma}^z_{3n-2} - \hat{\sigma}^z_{3n-1} \right) 
    \label{eq:Q_tot_3}\;,\\
    \hat{Q}^8_{tot} = \sum_{n=1}^N  \hat{Q}^8_n &= \frac{1}{2}\sum_{n=1}^N \sum_{ij} \hat{\phi}_n^{i\,\dagger} (\hat{\lambda}^8)_{ij} \,\hat{\phi}_n^{j} \nonumber\\
    &= \frac{1}{4\sqrt{3}}\sum_{n=1}^N \left(\hat{\sigma}^z_{3n-2} + \hat{\sigma}^z_{3n-1} - 2\hat{\sigma}^z_{3n} \right) \label{eq:Q_tot_8}\;,
\end{align}}%
where $\hat{\lambda}^3$ and $\hat{\lambda}^8$ are the diagonal Gell-Mann matrices. Using these conserved charges written in terms of Pauli matrices, it is straightforward to calculate the $\hat{K}$ matrix as it is completely diagonal.  Accordingly, the task reduces to evaluating the integral in Eq.~(\ref{eq:K_diagonal_general}) for each diagonal element of the projection operator.

\subsection{Reduction to irreducible representations}\label{sec:decomp_irreducible}
The diagonal matrices written in terms of Pauli operators (after the Jordan Wigner transformations) still form a representation of the underlying gauge groups. However, for a unit cell ($N=2$), this is not a fundamental representation nor is it irreducible. For example, for a unit cell of SU(2), $\hat{Q}_{tot}^z$ is a $2^{4} \times 2^{4}$ matrix. It is constructive to see the decomposition of this representation in terms of the irreducible representations as it illuminates the direct sum structure of the underlying Hilbert space and provides us information about the size of the singlet subspace within the full Hilbert space. In this section, we explain the decomposition of the representation used in Eq.~(\ref{eq:Qz_tot}), Eq.~(\ref{eq:Q_tot_3}), and Eq.~(\ref{eq:Q_tot_8}).\\

\textit{SU(2):---} We consider the  $\hat{Q}_{tot}^z$ operator for a single lattice site $(N=1)$, which in terms of the Pauli matrices is given by:
\begin{equation}
  \hat{Q}_{tot}^z (N=1) = \begin{pmatrix}
      0 & 0 & 0 & 0\\
      0 & 1/2 & 0 & 0\\
      0 & 0 & -1/2 & 0\\
      0 & 0 & 0 & 0
  \end{pmatrix}  \;.
\end{equation}
The eigenvalues $1/2, -1/2$ belong to the fundamental spin-1/2 representation of the SU(2) group. By writing the expressions for the non-diagonal charges $\hat{Q}_{tot}^x, \hat{Q}_{tot}^y$ for a single site (see Appendix \ref{app:unit-cell-hamiltonian}), it can be checked that the basis vectors $(1,0,0,0)$ and $(0,0,0,1)$ remain invariant under their operation, which corresponds to the zero eigenvalues of the $\hat{Q}_{tot}^z$ operator. This means that the two zero eigenvalues belong to the singlet representation. We can conclude that the representation ($D_Q$) for a single site charge operator consists of the fundamental representation ($D_{1/2}$) and two copies of the singlet representation ($D_0$):
\begin{equation}
    D_Q = D_{1/2} \, \oplus 2D_0\;.
\end{equation}
A single site can thus accommodate two singlet states, which in the strong coupling limit can be recognized as the vacuum and the antibaryon state. The $D_{1/2}$ representation belongs to the presence of single antiquark.

For two lattice sites or a unit cell $(N=2)$, the representation is a tensor product of two copies of the $D_Q$, which can be decomposed into direct sums. 
\begin{align}
    D_Q\, \otimes D_Q &= (D_{1/2} \, \oplus 2D_0) \otimes (D_{1/2} \, \oplus 2D_0) \nonumber\\
    & = D_1 \oplus 4D_{1/2} \,\oplus 5D_0\;,\label{eq:two-site-decomp-su2}
\end{align}
where the extra copy of the singlet originates from the direct sum decomposition $D_{1/2}\, \otimes D_{1/2} = D_{1}\,\oplus D_{0}$, i.e., the Hilbert space of two spin-1/2 particles consists of a triplet basis (or spin-1 representation) and a singlet basis. The decomposition in Eq.~(\ref{eq:two-site-decomp-su2}) can be continued for larger number of sites, where the direct sums decomposition can be determined by using Young's tableaux. 

From Eq.~(\ref{eq:two-site-decomp-su2}), we see that $N=2$ consists of five independent charge-singlet states, which creates the singlet subspace of the full Hilbert space. This is consistent with the number of independent singlet states one can construct on the unit cell. In the strong-coupling limit, these states are the vacuum, meson, baryon, antibaryon, and baryonium (see Appendix \ref{appsec:singlet-states}). The decomposition thus illuminates the size of the charge-singlet subspace for a gauge theory and one can iteratively determine the number of charge-singlet states for a given lattice size. An alternative way of finding the dimension of the singlet subspace is by finding the trace of the projection operator as mentioned in the discussion of Sec.~\ref{sec:general-projector} and is carried out explicitly in Sec.~\ref{sec:dimension_subspace}.

\textit{SU(3):---} The same decomposition can be performed for SU(3) LGT as well. However, for SU(3), the structure is more complicated due to presence of two diagonal charges $\hat{Q}_{tot}^3$ and $\hat{Q}_{tot}^8$. The representations $D(p,q)$ for SU(3) are denoted using two numbers $(p, q)$, corresponding to the eigenvalues of the two diagonal generators
\begin{equation}
    (Q_3)_{max} = \frac{p+q}{2}\,,\; (Q_8)_{max} = \frac{p-q}{2\sqrt{3}}\;.
\end{equation}
Here $((Q_3)_{max},(Q_8)_{max})$ denotes the coordinates of the right-most point in the $Q_3-Q_8$ plane, where $Q_3$, $Q_8$ represents the eigenvalues of the two diagonal operators. There are two inequivalent fundamental representations for SU(3) describing the particle ($D(1,0)$) and the antiparticle ($D(0,1)$).

To find the direct sum decomposition for SU(3) in the case of a single lattice site ($N=1$), we again take a look at the expression of the two diagonal operators in terms of the Pauli matrices, which is a $8\times 8$ matrix. The $(Q_3, Q_8)$ eigenvalue pairs are given by $(0,0)$, $(1/2,1/2\sqrt{3})$, $(1/2,-1/2\sqrt{3})$, $(0,-1/\sqrt{3})$, $(-1/2,-1/2\sqrt{3})$,$(-1/2,1/2\sqrt{3})$,$(0,1/\sqrt{3})$, $(0,0)$. Among these, the eigenvalue pairs $(1/2,1/2\sqrt{3})$, $(-1/2,1/2\sqrt{3})$,  $(0,-1/\sqrt{3})$ belong to the fundamental representation $D(1,0)$ and $(1/2,-1/2\sqrt{3})$, $(-1/2,-1/2\sqrt{3})$, $(0,1/\sqrt{3})$ belong to the fundamental representation $D(0,1)$. The two remaining $(0,0)$ eigenvalues belong to the two copies of the singlet representation 
\begin{equation}
    D_Q(N=1) = D(1,0)\, \oplus D(0,1)\, \oplus 2D(0,0)\,.
\end{equation}
Like the SU(2) case, the two singlet states for a single site can be recognized as the vacuum and antibaryon state, whereas the $D(0,1)$ and $D(1,0)$ representations include the antiquark and anti-diquark states which are not singlets.

For two sites $(N=2)$, the decomposition can be found by using Young's tableaux for the tensor product of two copies of $D_Q(N=1)$. One can show that it consists of six copies of the singlet representation, with two extra singlets coming from the result $D(1,0)\,\otimes D(0,1) = D(1,1)\,\oplus D(0,0)$, which leads to six orthogonal singlet basis states. This is consistent with the fact that a unit cell can host six singlet states, which in the strong coupling limit are vacuum, meson, baryon, antibaryon, tetraquark, and baryonium (see Appendix \ref{appsec:singlet-states}). Once again, we see that we gain information about the charge-singlet subspace and the structure of the Hilbert space from the matrix representation of the diagonal operators.

\subsection{Projection operator $\hat{K}$}\label{sec:proj_operator-K}
The projection operator for SU(2) and SU(3) LGTs (in (1+1)-D) can now be computed using the explicit expressions of the diagonal operators. The remaining element in Eq.~(\ref{eq:K_diagonal_general}) is the group-invariant measure of the integral $d\mu(\alpha_j)$, which depends on the parametrization $\alpha_j$ of the gauge group. Here we follow the same parametrization of the gauge groups SU(2) and SU(3) as described in~\cite{greiner2012quantum}
\begin{align}
 {\rm SU(2): \;}   \int d\mu(\alpha) &= \frac{1}{2\pi}\int_0^{4\pi} d\alpha\,\sin^2(\alpha/2)\;, \label{eq:SU2-parametrization}\\
 {\rm SU(3): \;} \int d\mu(a,b) &= \frac{4}{9\pi^2} \int_{-2\pi}^{2\pi} da \int_{-3\pi}^{3\pi} db\,\sin^2(a/2) \nonumber\\ &\sin^2(b/2+a/4)  \sin^2(b/2-a/4)\,. \label{eq:SU3-parametrization}
\end{align}
For the parametrized measures above, the parametrized form of a generic group element $\hat{U}_d(\boldsymbol{\alpha})$ belonging to the Cartan subgroup also needs to be specified. For SU(2) and SU(3), the corresponding group elements are parametrized as $\hat{U}_d(\alpha) = e^{i\alpha\hat{Q}^z_{tot}}$ and $ \hat{U}_d(a,b) = e^{i(a\hat{Q}_{tot}^3 + 2b\hat{Q}_{tot}^8/\sqrt{3})}$, respectively. We will use these parametrizations to evaluate the projection operator $\hat{K}$.    \\

\textit{SU(2):---} For SU(2), we can now evaluate the integral in Eq.~(\ref{eq:K_diagonal_general}). Using the explicit expression in Eq.~(\ref{eq:Qz_tot}), we can evaluate the integral of the diagonal terms of $e^{i\alpha \hat{Q}_{tot}^z}$, which are of the form $e^{i\alpha k/2}$, where $k$ are non-negative integers. This stems from the fact that the eigenvalues for all irreducible representations of SU(2) are either integers or half integers. So, evaluating Eq.~(\ref{eq:K_diagonal_general}) becomes equivalent to the following:
\begin{equation}
    \frac{1}{2\pi}\int_0^{4\pi} d\alpha\,\sin^2(\alpha/2)\, e^{i\alpha k/2} = \begin{cases}
        1 \qquad\qquad\text{if } k = 0\\
        -1/2 \;\qquad\text{if } k = \pm 2\\
        0 \qquad\qquad\text{otherwise }
    \end{cases}.
\end{equation}
This result also gives us an insight about the dimension of the charge-singlet subspace. Each zero eigenvalue yields $1$ in the diagonal entry. However, the zero eigenvalue can either belong to a singlet representation or it can belong to a $D_j$ representation where $j$ is a positive integer. For SU(2), in each $D_j$, we have eigenvalues $1$ and $-1$ which appear in the diagonal, corresponding to $k=\pm 2$ in the expression above. So, the number of times $-1/2$ appears in the diagonal entries of the operator $\hat{K}$ is twice the number of irreducible representations $D_j$ where zero appears as an eigenvalue. So, counting the number of times 1 appears in the diagonal and subtracting half the number of times $-1/2$ appears in the diagonal will also give us the dimension of the charge-singlet subspace. The counting of charge-singlet representations then just reduces to the trace of the projection operator, which is used in Sec.~\ref{sec:dimension_subspace}.

Since we want to measure the $\hat{K}$ operator on a quantum device, we want to study the Pauli decomposition of the operator. It is evident from the expression of $\hat{Q}_{tot}^z$ that only $\hat{\sigma}^z$ and identity operators will appear in the decomposition as $\hat{Q}_{tot}^z$ is a diagonal matrix. So, measuring $\hat{K}$ on a device can be done using $z-$basis measurements only. However, we want to still count the number of separate Pauli strings that appear in the decomposition of $\hat{K}$. This is particularly important to know in the case of measuring the expectation value of a  non-diagonal observable using Eq.~(\ref{eq:exp-val-K}). In that case, we need to measure the expectation $\langle\hat{O}\hat{K}\rangle$ on the non-charge-singlet state prepared on a quantum device. The number of Pauli strings present in the composite operator $\hat{O}\hat{K}$ depends on the Pauli string decomposition of both $\hat{K}$ and $\hat{O}$.

The integrand in Eq.~(\ref{eq:K_diagonal_general}) contains an exponential $e^{i\alpha \hat{Q}_{tot}^z}$ which can be written in terms of Pauli operators
\begin{align}
    e^{i\alpha \hat{Q}_{tot}^z} &= \prod_{n=1}^N e^{i\frac{\alpha}{4}\hat{\sigma}^z_{2n-1}} e^{-i\frac{\alpha}{4}\hat{\sigma}^z_{2n} } \nonumber\\
    &= \prod_{n=1}^N [\cos(\alpha/4)\, I + i \sin(\alpha/4)\,\hat{\sigma}^z_{2n-1}]\times\nonumber\\
    & \qquad\qquad[\cos(\alpha/4)\, I - i \sin(\alpha/4)\,\hat{\sigma}^z_{2n}]\,.
\end{align}
Expanding this product in terms of the Pauli operators $\hat{\sigma}^z$ will resemble the binomial expansion of power $2N$. This means the $\alpha$-dependent part of the coefficients of the Pauli strings of different lengths will be of the form $\cos^m(\alpha/4) \sin^{2N-m}(\alpha/4)$. The number of such terms (each with the same $\alpha$-dependent coefficient but associated with different Pauli strings of same length) is given by the corresponding binomial coefficients $^mC_{2N-m}$, where $C$ denotes the combination symbol. Using properties of sine and cosine functions, it is straightforward to show that in the integral Eq.~(\ref{eq:K_diagonal_general}) only even $m$ terms survive
\begin{align}
    \frac{1}{2\pi}\int_0^{4\pi} d\alpha\sin^2(\alpha/2) \cos^m(\alpha/4) &\sin^{2N-m}(\alpha/4) = 0 \,,
    \nonumber\\&\qquad\;
    \text{$\forall$ odd $m$}\;.
\end{align}
The total number of even terms with different Pauli strings is exactly half of the total $2^{2N}$ terms. So, the Pauli string decomposition of $\hat{K}$ contains $2^N$ Pauli strings, though all of them are diagonal.
For a non-diagonal observable like the Hamiltonian, the composite operator $\hat{O}\hat{K}$ can thus contain exponentially many Pauli strings which are not diagonal. A convenient workaround is to use a Monte-Carlo sampling method to evaluate the integral in Eq.~(\ref{eq:K_diagonal_general}) instead of using the closed form expressions found here. This is elaborated in Sec.~\ref{sec:GI-entropy} in greater detail.\\

\textit{SU(3):---} Most discussions in the subsection for SU(2) can be generalized rather straightforwardly to SU(3), albeit being algebraically more complicated. The presence of two diagonal charges leads to the parametrization in terms of two parameters $a, b$ as given in Eq.~(\ref{eq:SU3-parametrization}). As mentioned before, a generic group element of the Cartan subalgebra is given by
\begin{equation}
    \hat{U}_d(a,b) = e^{i(a\hat{Q}_{tot}^3 + 2b\hat{Q}_{tot}^8/\sqrt{3})}\;. \label{eq:integrand-su3}
\end{equation}
The eigenvalues of the integrand are of the form $e^{i(ma/2+nb/3)}$, where $m$ and $n$ are integers, which depend on the representations $D(p,q)$ that appears in the direct sum decomposition for a given number of lattice sites. 
The double integral in Eq.~(\ref{eq:SU3-parametrization}) can now be performed analytically for all the diagonal elements in Eq.~(\ref{eq:integrand-su3}). As in the case of SU(2), all zero eigenvalues contribute a diagonal entry of 1. This is compensated by other diagonal entries for each zero not belonging to the singlet representation, yielding the trace of $\hat{K}$ as the dimension of the singlet subspace. 

Similar to SU(2), we can also count the number of diagonal Pauli strings that appear in the decomposition of the projection operator $\hat{K}$. Expanding the exponent of Eq.~(\ref{eq:integrand-su3}) for a single site results in
\begin{equation}
    a\hat{Q}_{n}^3 + \frac{2b}{\sqrt{3}}\hat{Q}_{n}^8 = \left(\frac{a}{4}+\frac{b}{6}\right)\hat{\sigma}_{3n-2}^z + \left(\frac{a}{4}-\frac{b}{6}\right)\hat{\sigma}_{3n-1}^z - \frac{b}{3} \hat{\sigma}_{3n}^z\;.
\end{equation}
The $a,b$ dependent part of the coefficients for Pauli strings are then of the form $\cos^m\lambda\sin^{3N-m}\gamma$, where $N$ is the number of spatial sites on the lattice, and $\lambda, \gamma\in\{a/4\pm b/6,-b/3\}$. As in the case of SU(2), the number of terms with the coefficients mentioned here is $^mC_{3N-m}$, which follows the binomial coefficients with degree $3N$. Only half of these coefficients (with even $m$) yields a non-zero contribution to the expansion, leaving $2^{3N/2}$ Pauli strings (total number of even terms) in the decomposition of diagonal operator $\hat{K}$.



\section{Application of the projection method to thermal states}\label{sec:thermodynamics-section}
In Secs.~\ref{sec:theory-framework} and \ref{sec:diagonal-projector}, we outlined how to construct the projection operator explicitly and apply it to compute observables within the charge-singlet subspace. In this section, we illustrate how this approach was implemented in our work \cite{than2024phase} to extract expectation values of observables in lattice gauge theory simulations at finite temperature. Complementary to the evaluation of a fermionic observable in \cite{than2024phase}, we turn our attention here to reading out properties of the gauge field, more specifically the expectation value of the electric field Hamiltonian.
We further extend the utility of this method for thermal states by introducing a new technique for accessing thermodynamic quantities like entropy, for charge-singlet density matrices. In this context, we also demonstrate how classical simulation techniques—such as tensor networks—can leverage this framework, broadening its applicability to both classical and quantum simulations.

\subsection{Observable measurement for thermal states in lattice gauge theories}\label{sec:thermal-states}
Here, we briefly explain the use of the charge-singlet measurement technique for the preparation of thermal states for SU(2) and SU(3) non-Abelian gauge theories with matter in (1+1)-D. One major challenge in preparing thermal states for gauge theories with the color-neutrality condition is that the prepared Gibbs state must be a probabilistic mixture of charge-singlet states that resides within a subspace of the full Hilbert space.
Enforcing the color-neutrality constraint in variational algorithms typically involves incorporating penalty terms into the cost function or designing symmetry-protecting circuits. However, these strategies become significantly more challenging for non-Abelian gauge theories, where multiple non-commuting conserved charges are present. The difficulty is further compounded in the context of thermal state preparation, which requires generating a probabilistic mixture of charge-singlet states. This entails not only constructing circuits capable of preparing individual singlet states but also ensuring they are mixed with the correct Boltzmann weights. Achieving both objectives simultaneously is considerably harder than preparing a single charge-singlet state.

In \cite{than2024phase}, we introduced an alternative approach for incorporating the effects of charge-singlet constraints at finite temperature using the projection-based charge-singlet measurement (CSM) technique. A key advantage of this method is that it eliminates the need to enforce the constraints during the state preparation stage. This flexibility allows us to choose any efficient thermal state preparation protocol (not limited to protocols designed only for gauge theories anymore), providing us more freedom in circuit designing and variational ansatz construction. In our work, we adopted the Variational Quantum Thermalizer (VQT) algorithm \cite{consiglio2023variational} to prepare thermal states. This protocol uses an ancilla register to sample bitstrings from a parameterized probability distribution, which is variationally optimized (see Appendix \ref{appsec:VQT-circuit}). A system register with parameterized gates is then initialized with the sampled bitstring and is tasked with generating the eigenbasis of the density matrix. By minimizing the free energy as the cost function, the system converges toward the Gibbs state of the full Hilbert space, which, in the case of our lattice gauge theory, belongs to a reducible representation. We can now use the projection method to measure the expectation value of an observable $\hat{O}$ restricted to the singlet representation only. From Eq.~(\ref{eq:exp-val-K}), this involves measuring both $\hat{O}\hat{K}$ and $\hat{K}$ on the reducible density matrix $\hat{\rho}$ produced at the end of the VQT protocol. Once again, we note that even though we never explicitly construct the charge-singlet density matrix, our method still yields the correct singlet-sector expectation value, which is usually the main goal of the experiment.

Here, we use Eq.~(\ref{eq:exp-val-K}) to determine the expectation value of a gauge-field dependent observable $\langle\hat{H}_{el}\rangle_{0}$ on the singlet subspace. As a concrete example, we consider a unit cell of SU(2) LGT in (1+1)-D at a finite temperature $T=0.5$. The explicit forms of the electric field Hamiltonian $\hat{H}_{el}$ and the full Hamiltonian $\hat{H}$ are given in Appendix~\ref{app:unit-cell-hamiltonian}. The thermal state $\hat{\rho}$ is prepared using the noiseless variational protocol VQT as explained in Appendix \ref{appsec:VQT-circuit}. Since $\hat{H}_{el}$ is a diagonal observable, all measurements are performed in the computational basis to obtain the expectation values of $\hat{H}_{el}\hat{K}$ and $\hat{K}$.

\begin{figure}[!ht] %
    \centering
    \includegraphics[width=1.0\linewidth]{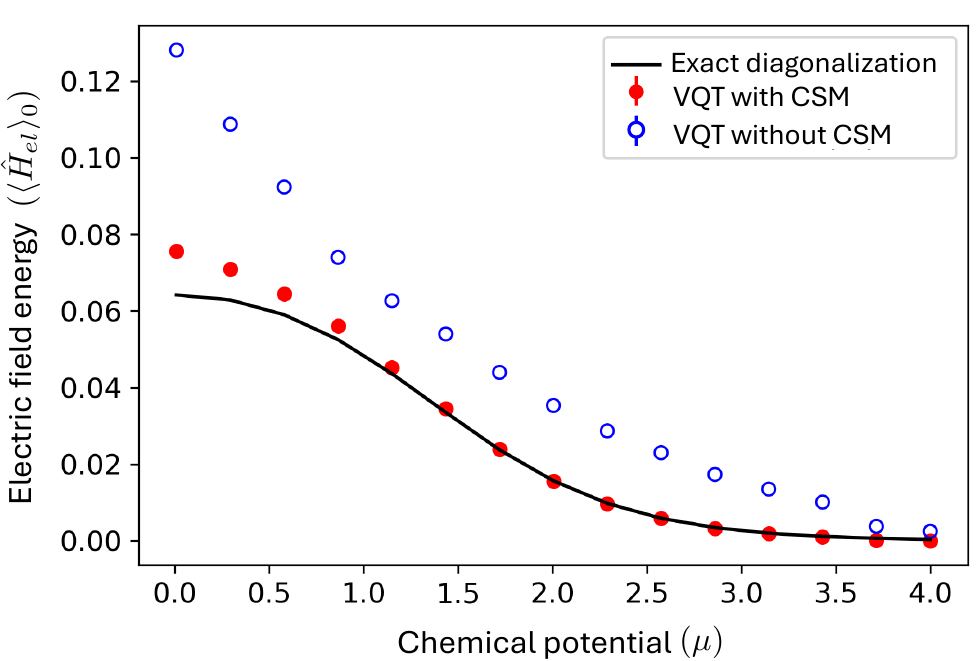} 
    \caption{\textbf{Electric field versus chemical potential for a SU(2) unit cell.} The VQT protocol as introduced in \cite{than2024phase} is employed to numerically simulate thermal states for a SU(2) unit cell at a fixed temperature $T=0.5$, with Hamiltonian parameters $m=g^2=0.5$. As the chemical potential $\mu$ is varied, the expectation value of the electric field Hamiltonian  $\langle \hat{H}_{el}\rangle_0$ changes accordingly. Our VQT protocol combined with the projection technique described in Eq.~(\ref{eq:exp-val-K}) yields results that closely agree with those from exact diagonalization. In contrast, the expectation value calculated from the reducible density matrix $\hat{\rho}$ without the projection formula leads to inaccurate results due to $\hat{\rho}$ being a mixture of energy eigenstates from different irreducible representations. For each value of $\mu$, the variational optimization is repeated across five independent trials with random initializations. The error bars shown represent the standard deviation of the optimized electric field expectation values across these trials, although they are too small to be visible in the plot.}
    \label{fig:Electric-field-SU2}
\end{figure}
%

In Fig.~\ref{fig:Electric-field-SU2} we see that the desired expectation value of the electric field Hamiltonian is accurately captured by the projection method without actually preparing the charge-singlet density matrix. For comparison, we show that ${\rm Tr}(\hat{\rho}\hat{H}_{el})$ is not the correct expectation value as $\hat{\rho}$ does not belong to the charge-singlet subspace. The slight deviation of the values obtained from VQT with the projection method (shown with red spherical markers) on the left side of the plot is consistent with the results obtained in \cite{than2024phase}. This is a consequence of the fact that at low chemical potential values, more states are present in the thermal mixture and the VQT may fail to capture the correct Boltzmann weight of some of these states, leading to a slight deviation from the exact value. Therefore, this is not a failure of the projection method, but rather the artifact of using a simple VQT ansatz. From Fig.~\ref{fig:Electric-field-SU2}, we then infer that Eq.~(\ref{eq:exp-val-K}) can be successfully used to determine any observable expectation value for a thermal state prepared on the full Hilbert space.

\subsection{Charge-singlet entropy}\label{sec:GI-entropy}
In this section, we demonstrate the broader utility of the CSM framework for thermal states, extending beyond the applications presented in \cite{than2024phase} and Sec.~\ref{sec:thermal-states}. In particular, we show that this method enables the calculation of the von Neumann entropy of the equilibrium density matrix, projected onto the charge-singlet subspace at finite temperature. Accessing the von Neumann entropy—a fundamental thermodynamic quantity—in the context of lattice gauge theories could provide new avenues for numerically or experimentally exploring their connection with quantum thermodynamics, a recently developed field of study \cite{majidy2023noncommuting,davoudi2024quantum}.

The expression of the charge-singlet entropy $S_0$ can be derived from the relation 
\begin{equation}
    \langle K \rangle = {\rm Tr}(\hat{\rho} \hat{K}) = \frac{{\rm Tr}(e^{-\beta\hat{H}}\hat{K})}{Z} = Z_0/Z \;, \label{eq:z0-z-K-relation}
\end{equation}
where $Z_0 = {\rm Tr}(e^{-\beta \hat{H}}\hat{K})$ is the charge-singlet partition function and $Z = {\rm Tr}(e^{-\beta\hat{H}})$ is the full partition function for the reducible representation.
The charge-singlet free energy can be written in terms of the charge-singlet partition function  $F_0 = -T \ln Z_0 = \langle \hat{H} \rangle_0 - T S_0$ and the non-restricted (reducible) free energy $F = -T \ln Z = \langle H\rangle - TS$. The  entropy $S_0$ is then given by the expression
\begin{equation}
    S_0 = S + \ln\langle \hat{K} \rangle + \frac{1}{T\langle \hat{K} \rangle} \left(\langle \hat{H} \hat{K} \rangle - \langle \hat{H} \rangle \langle \hat{K} \rangle\right)\;, \label{eq:gi_entropy}
\end{equation}
where we have used $\langle \hat{H} \rangle_0 = \langle \hat{H} \hat{K} \rangle/ \langle \hat{K} \rangle$. The averages in Eq.~(\ref{eq:gi_entropy}) are measured with respect to the reducible density matrix $\hat{\rho}$.
In calculating $S_0$, we thus need to measure operators $\hat{H}$, $\hat{K}$, and $\hat{H}\hat{K}$. Out of these operators, $\hat{H}$ contains a polynomial number of Pauli strings, and $\hat{K}$ is completely diagonal. However, $\langle \hat{H} \hat{K} \rangle$ can in general contain exponentially many non-diagonal Pauli strings (due to $\hat{K}$ having exponentially many diagonal Pauli strings), which can make the evaluation of the entropy in Eq.~(\ref{eq:gi_entropy}) resource-expensive on a quantum computer.

\begin{figure}[ht]
    \centering
    \includegraphics[width=1.0\linewidth]{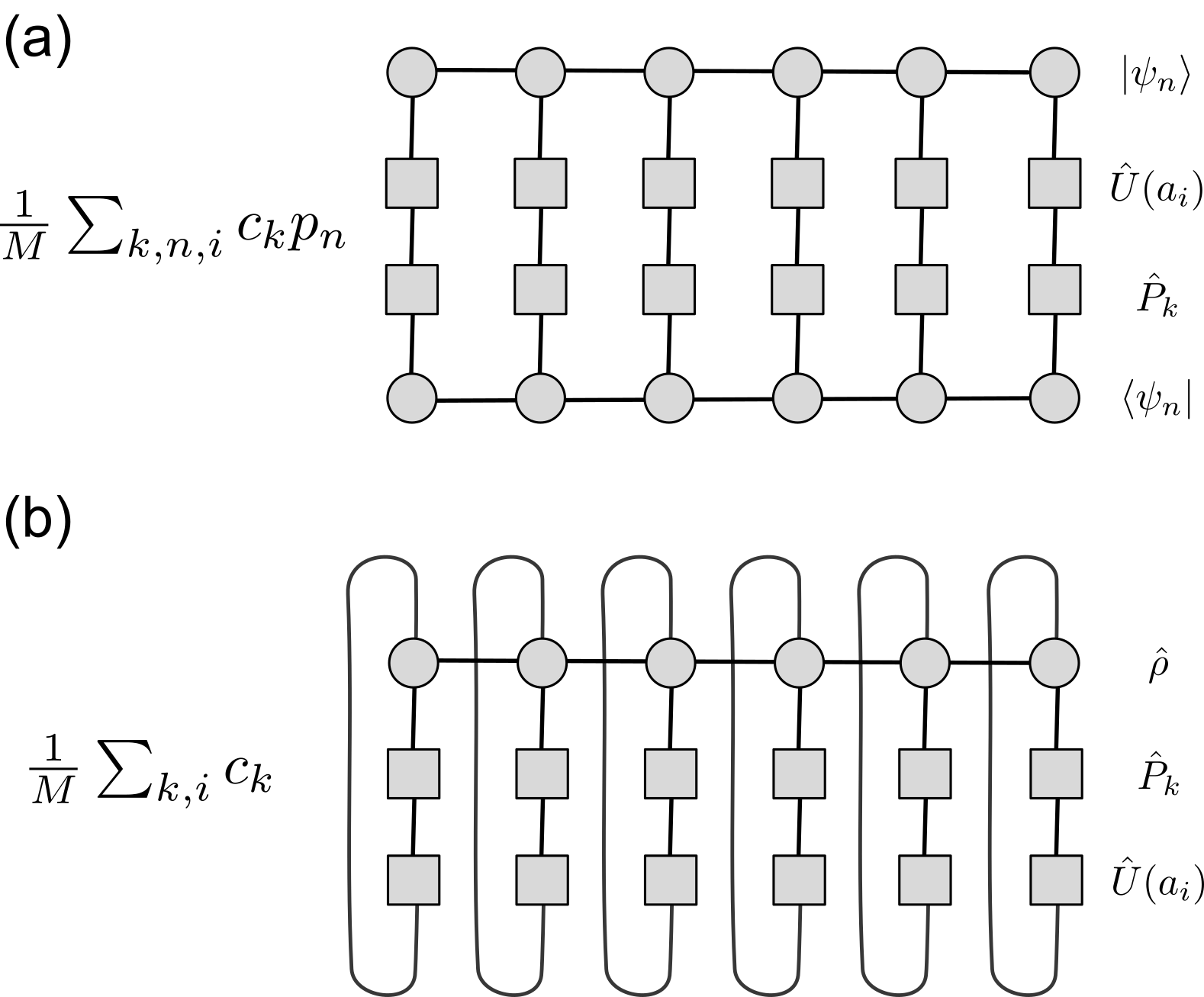}
    \caption{\textbf{Evaluation of charge-singlet entropy in a classical simulation.} (a) Matrix product states can be used to evaluate quantities like $\langle \hat{H}\hat{K}\rangle$. In this case, the reducible density operator $\hat{\rho}$ is expanded using $\hat{\rho} = \sum_n p_n |\psi_n\rangle\langle\psi_n|$, where $|\psi_n\rangle$ is written in the MPS form and $\hat{H} = \sum_k c_k \hat{P}_k$, where $\hat{P}_k$ are Pauli strings. $a_i$ are Monte Carlo samples for the single-qubit operators, and $M$ is the number of Monte Carlo samples used to evaluate the expectation value. (b) Instead of using a linear superposition of MPS, one can also use MPO to represent the reducible density matrix $\hat{\rho}$ for evaluating Eq.~(\ref{eq:MC_integral}).}
    \label{fig:MPS_entropy}
\end{figure}

Nevertheless, as noted in the introduction, the projection technique is also applicable to classical simulation methods such as tensor networks (TNs). Given a TN representation of the full density matrix $\hat{\rho}$, one can efficiently compute the expectation values of $\hat{H}\hat{K}$ and $\hat{K}$. In (1+1) dimensions, matrix product operators (MPOs) have been successfully used to represent mixed-state density matrices and can be variationally optimized to  approximate thermal states \cite{guth2020efficient,cui2015variational,banuls2015thermal,tang2020continuous}. While the representation of thermal states using TNs remains an active area of research, our method for measuring entropy via the projection technique is expected to remain applicable to future protocols—as long as the evaluation of tensor products of single-site operators remains efficient. In this section, we used a TN to emulate the optimized variational circuit for a SU(2) unit cell employed in Sec.~\ref{sec:thermal-states}, yielding a TN representation of the reducible density matrix \cite{zhang2023tensorcircuit}. Alternatively, current and emerging TN-based methods may be used to variationally create a classical representation of the density matrix $\hat{\rho}$.   

Given such a TN representation of $\hat{\rho}$ the evaluation of entropy can be made efficient using Monte-Carlo sampling, instead of using an explicit expression for $\hat{K}$. To understand this, we use the definition of the operator $\hat{K}$
\begin{equation}
    {\rm Tr}(\hat{\rho} \hat{O}\hat{K}) = \int d\mu(\boldsymbol{a})\; {\rm Tr}(\hat{\rho} \hat{O}\hat{U}_d(\boldsymbol{a}))\;, \label{eq:MC_integral}
\end{equation}
where $U_d(\boldsymbol{a}) = e^{i\boldsymbol{a}\cdot\mathbf{\hat{Q}}^{diag}_{tot}}$ is a group element that can be defined in terms of diagonal conserved charges only. Since the diagonal charge operators $\mathbf{\hat{Q}}_{tot}^{diag} = \sum_n \mathbf{\hat{Q}}_{n}^{diag} $ are separable at each lattice vertex, the operator $\hat{U}(\boldsymbol{a})$ can be written as a tensor product of single qubit operators. For example, in SU(2) LGT, the operator $\hat{U}(a) = e^{ia \hat{Q}_{tot}^z} = \bigotimes_{k=1}^{2N} e^{ia(-1)^{k+1}\hat{\sigma}^z_k/4}$, where $N$ is the total number of lattice sites and we have used a compact form of Eq.~(\ref{eq:Qz_tot}). Due to the separable form of the operator  $\hat{U}(\mathbf{a})$, this can be applied efficiently to a matrix product operator or a matrix product state. The trace in Eq.~(\ref{eq:MC_integral}) can be evaluated by efficient contractions of the tensor network (see Fig.~\ref{fig:MPS_entropy}). The integral in Eq.~(\ref{eq:MC_integral}) is then replaced by a sum over Monte-Carlo samples where $\mathbf{a}$ is sampled from a distribution that follows the group measure $d\mu(\mathbf{a})$.

\begin{figure}[ht]
    \centering
    \includegraphics[width=1.0\linewidth]{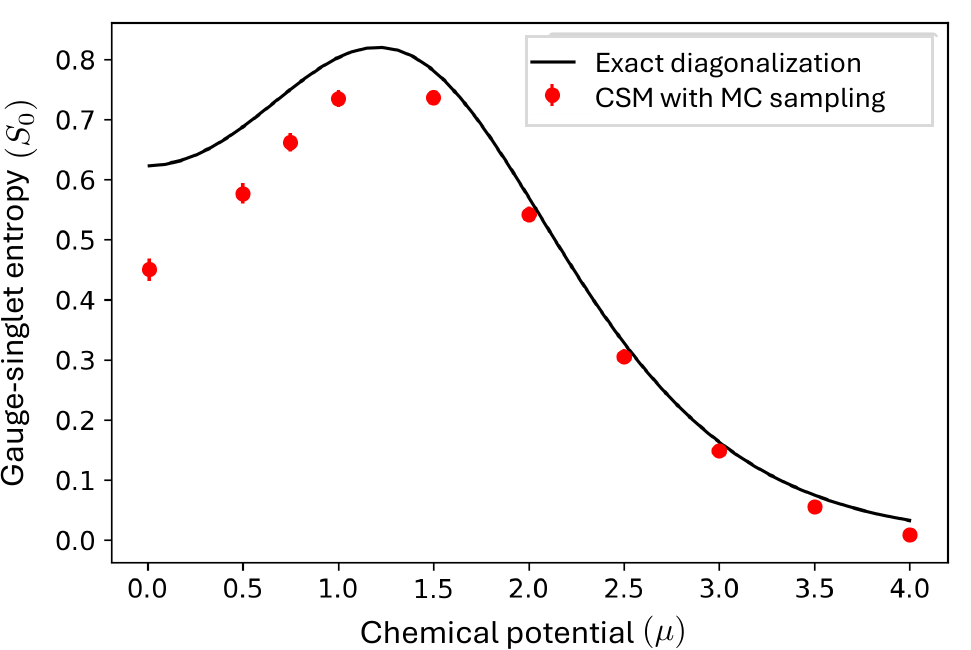}
    \caption{\textbf{Charge-singlet entropy with Monte-Carlo sampling.} The entropy of the charge-singlet density matrix for a SU(2) unit cell at finite temperature ($T=0.5$) is computed using Eq.(\ref{eq:gi_entropy}) within the tensor network (TN) framework. The TN is implemented as a matrix product operator (MPO) constructed from the optimized reducible density matrix $\hat{\rho}$ obtained at the conclusion of the VQT protocol described in Sec.~\ref{sec:thermal-states}. Once the MPO is constructed, the quantities $\langle \hat{K}\rangle$ and $\langle \hat{H}\hat{K}\rangle$ in Eq.~(\ref{eq:gi_entropy}) are evaluated using Monte Carlo (MC) sampling (2000 samples for each value of $\mu$) combined with tensor contractions, as illustrated in Fig.~\ref{fig:MPS_entropy}. The charge-singlet entropy, computed via the projection operator in Eq.~(\ref{eq:gi_entropy}) and MC sampling, shows good quantitative agreement with results from exact diagonalization. The  discrepancies observed at lower chemical potentials arise from limitations in the expressiveness of the variational circuit used in the VQT protocol, rather than from sampling error. The error bars represent the standard deviation computed over five independent trials, each using a different random seed for the Monte Carlo sampling.}
    \label{fig:gi_entropy}
\end{figure}

As a concrete example for the evaluation of the entropy, we show here the evaluation of entropy for a single unit cell for SU$(2)$ lattice gauge theory. In SU(2) LGT, there is only one diagonal conserved charge $\hat{Q}_{tot}^z$, that can be compactly written as (from Eq.~(\ref{eq:Qz_tot})) 
\begin{equation}
    \hat{Q}_{tot}^z = \frac{1}{4}\sum_{k=1}^{2N} (-1)^{k+1} \hat{\sigma}_k^z\;.
\end{equation}
The measure of the SU(2) group for this particular parametrization is already given in Eq.~(\ref{eq:SU2-parametrization}). To evaluate the average $\langle\hat{H}\hat{K}\rangle$, and $\langle\hat{K}\rangle$, we draw $2000$ Monte Carlo samples for $a$, that follows the distribution $d\mu(a)$. 

We then use either a matrix product state (MPS) or a matrix product operator (MPO) to compute the expectation values efficiently, leveraging the contraction schemes inherent to tensor networks. The non-charge-singlet entropy 
$S$ can be determined either analytically for simpler circuits (as in \cite{than2024phase}) or by sampling the ancilla register in the variational ansatz in the $z$-basis. To construct the TN representation of $\hat{\rho}$, we used the optimized parameters obtained from the VQT algorithm employed in Sec.~\ref{sec:thermal-states} at $T = 0.5$. We then executed a tensor circuit simulation to translate the corresponding optimized quantum circuit output into an MPS or MPO, using the Python package introduced in \cite{zhang2023tensorcircuit}.
Subsequently, we applied Monte Carlo sampling to evaluate the charge-singlet entropy as discussed in this section, which shows excellent agreement with exact diagonalization results (see Fig.~\ref{fig:gi_entropy}). The observed deviation in entropy at low chemical potential is not due to the Monte Carlo sampling method but arises as a consequence of the limitations of the simple VQT circuit, which reflects in the MPO representation of the optimized $\hat{\rho}$. The method described here is scalable due to the sampling process and can be carried out efficiently in classical simulations.
It is worth mentioning that in our simulation, due to a short circuit depth, the MPS or MPO representation incur negligible approximation error. For a larger systems and deeper circuits, the bond dimension needs to be truncated, which may introduce approximation errors. 

\section{Noise mitigation for gauge theories}\label{sec:noise-mitigation}
In Secs.~\ref{sec:thermal-states} and \ref{sec:GI-entropy}, we described how the CSM technique can be used in a variational protocol that does not preserve the total color charge. In the case of the thermal state preparation protocol, the initial states were not charge singlets, and the VQE ansatz was not color-symmetry preserving. Still, by using the projection technique, we recovered expectation values of observables measured on the singlet subspace using the reducible density matrix. However, the application of the projection method is not restricted to symmetry-violating ansatzes only. It can also be used as a practical noise-mitigation tool for quantum simulations where the charge-singlet property is protected by virtue of the operations performed on the system. In particular, the CSM technique can be used to get a better estimate of observables in noisy implementation of symmetry-preserving circuits. In this section, we illustrate a few such examples where the CSM approach can be used, which includes time evolution (Sec.~\ref{sec:noisy-time-evolve}) and ground state preparation (Sec.~\ref{sec:ground-state-prep}). In the context of time evolution in Sec.~\ref{sec:noisy-time-evolve}, we also demonstrate how CSMs can be used to quantify the depolarizing error corresponding to two-qubit gates in a LGT simulation.

\subsection{Time evolution}\label{sec:noisy-time-evolve}
Traditional Lagrangian-based approaches of lattice gauge theories encounter significant difficulties in predicting dynamical properties of a system due to the time coordinate being Euclideanized. In Hamiltonian based approaches, time is a free parameter and we can study dynamics of quantum systems with an underlying gauge symmetry. So, quantum computers can offer a natural advantage in studying time evolution of LGTs. 

In general, time evolution in quantum computers is performed using trotterization. Given an initial state $\ket{\psi}$, the time evolved state $e^{-i\hat{H}t}\ket{\psi}$ is determined approximately by performing the trotterized operation
\begin{equation}
    \ket{\psi_t} \equiv e^{-i\hat{H}t}\ket{\psi} \approx \left(\prod_j e^{-i\hat{h}_j\delta t}\right)^{N_t} \ket{\psi}\;,
    \label{eq:trotter_def}
\end{equation}
where $\delta t$ is the trotter step size, $N_t\delta t = t$ and $\hat{H} = \sum_j \hat{h}_j$. Here $\hat{h}_j$s are the parts of the Hamiltonian that do not commute with each other. If the trotter errors are small enough and $\ket{\psi}$ is a charge-singlet state, $\ket{\psi_t}$ should remain a charge-singlet state as the time-evolution operator $e^{-i\hat{H}t}$ preserves the total color charge.

For a given Hamiltonian $\hat{H}$, the operations $e^{-i\hat{h}_j\delta t}$ are performed using quantum gates. Since $\hat{h}_j$ can involve many-body terms, the operation will generally involve both single- and two-qubit rotation gates. For a perfect implementation of these gates, the final state will still be a charge-singlet (given the trotter errors are small enough). However, for current quantum hardware, the gates are noisy and even for high two-qubit gate fidelities, longer time evolutions (resulting in deeper circuits) can induce significant errors, leading to a noisy final state which is not a charge-singlet anymore. In fact, for a very deep circuit, the resulting state would asymptotically reach the maximally mixed state.

The CSM technique can also be used in this noisy time-evolution case, where the existence of noise is the main reason behind the violation of color-neutrality. In order to measure an operator $\hat{O}$ expectation value on a time evolved state $\ket{\psi_t}$, we use 
\begin{equation}
    \langle \hat{O}\rangle(t) = \frac{{\rm Tr}(\hat{\rho}(t)\hat{O}\hat{K})}{{\rm Tr}(\hat{\rho}(t)\hat{K})}\;, \label{eq:time_evol_proj}
\end{equation}
where $\hat{\rho}(t)$ is the noisy time-evolved state prepared on the device. The projected density matrix $\hat{\rho}_0 (t)$ is closer to the desired pure state \cite{ballini2024symmetry}. To show this, we note that the $\hat{\rho}_0 (t)$ is given in terms of $\hat{\rho}(t)$ by Eq.~(\ref{eq:singlet-density-projection-od}) and the fidelity between the prepared state $\hat{\rho}(t)$ with the desired state $\ket{\psi_t}$ is given by ${\rm Tr}(\hat{\rho}_0\ket{\psi_t}\bra{\psi_t})$. We obtain
\begin{equation}
   {\rm Tr}(\hat{\rho}_0\ket{\psi_t}\bra{\psi_t}) =  \frac{{\rm Tr}(\hat{\rho}(t) \ket{\psi_t}\bra{\psi_t}\hat{K})}{{\rm Tr}(\hat{\rho}(t)\hat{K})}\ge {\rm Tr}(\hat{\rho}(t)\ket{\psi_t}\bra{\psi_t})\;,
\end{equation}
where we have used the fact that within the trace sign $\hat{P}_0$ can be replaced with $\hat{K}$ and $\hat{K}\ket{\psi_t}=\ket{\psi_t}$ since $\ket{\psi_t}$ is a charge-singlet state. We have also used ${\rm Tr}(\hat{\rho}(t)\hat{K})< 1$ as $\hat{K}$ is the projection operator. This explains that applying the projection method to a noisy trotterized circuit performing time evolution can yield improved observable estimates. A related method has been used in \cite{ballini2024symmetry} for symmetry verification of time evolution of a discrete non-Abelian $D_3$ gauge theory. In this article, we extend the application to gauge groups relevant to the Standard Model of particle physics, and to realistic, device-aware noise models for trapped-ion systems, as we demonstrate below.

\textit{Example:}--- As a demonstration of using the CSM technique for time evolution, we take the specific case of a single unit cell for SU(2) non-Abelian lattice gauge theory with fermionic matter in (1+1)-D as described in Sec.~\ref{sec:thermodynamics-section} and Appendix \ref{appsec:su2-Ham}. We consider a time evolution starting from the strong-coupling
vacuum state (which is a charge-singlet). Since the full Hamiltonian is used for this time evolution on the strong-coupling vacuum state, particle-antiparticle pairs are created and annihilated as time progresses. The expectation value of the mass Hamiltonian as a function of time effectively captures this dynamics of pair creation and annihilation.

\begin{figure}[h]
    \centering
    \includegraphics[width=1.0\linewidth]{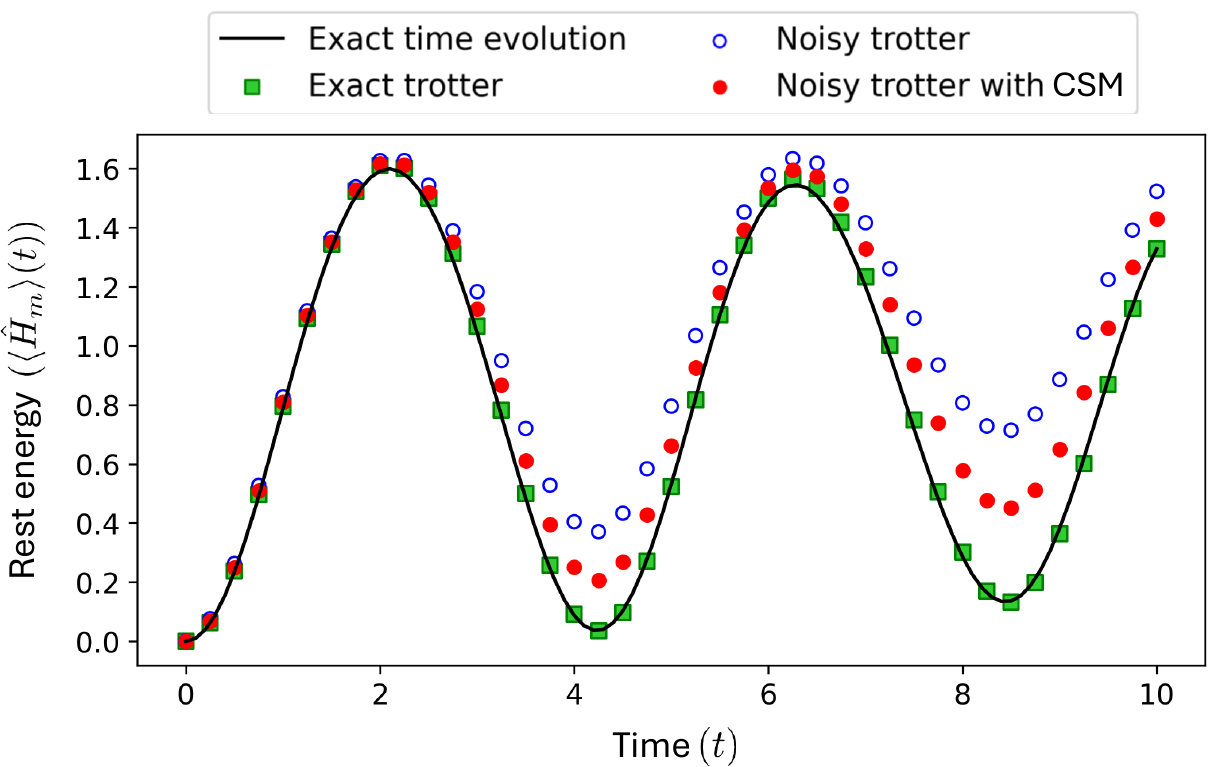}
    \caption{\textbf{Noisy time evolution of a SU(2) unit cell with fermionic matter in (1+1)-D.} Starting from the strong-coupling vacuum state, the system is evolved in time till $t=10.0$ using the full Hamiltonian with $m=g^2=0.5$. The trotter step size is chosen to be $\delta t = 0.25$. Noisy time evolution is simulated using a noise model with a two-qubit gate fidelity of $\sim99.9\%$, implemented via a depolarizing channel applied after each two-qubit gate.}
    \label{fig:Time-evolution-mitigated}
\end{figure}

The Hamiltonian for the unit cell can be separated into two non-commuting parts: one involving the kinetic non-diagonal part of the Hamiltonian and the other involving the diagonal terms. More details of the SU(2) Hamiltonian and the parameters are given in Appendix~\ref{app:unit-cell-hamiltonian}. Time evolution can then be performed using trotterization, with the many-body terms appearing in the kinetic and electric part of the Hamiltonian expressed in terms of single and two-qubit rotation gates. Without any circuit optimization, each trotter step includes 13 CNOT gates (see Appendix~\ref{app:circuits}). In Fig.~\ref{fig:Time-evolution-mitigated}, the trotter error can be seen to be negligible when the CNOT gates are assumed to be perfect. However, in a near-term quantum device, noise in two-qubit gates is usually the dominant error and controls the quality of the simulation. As a noise model in this particular example, we considered each CNOT gate followed by a local (two-qubit) depolarizing channel of strength $\lambda_d = 0.001$ (see Appendix \ref{appsec:noise-model}. Even with a CNOT gate with $\sim 99.9\%$ fidelity, we notice significant deviation (blue circles in Fig.~\ref{fig:Time-evolution-mitigated}) with large number of trotter steps as expected. This deviation is strictly coming from the gate errors and we can correct for the part that violates the color-neutrality condition of the resulting state using Eq.~(\ref{eq:time_evol_proj}). By virtue of the projection method, we observe that using Eq.~(\ref{eq:time_evol_proj}) to measure $\langle \hat{H}_{m}\rangle$ gives us an estimate of the expectation value (red circles in Fig.~\ref{fig:Time-evolution-mitigated}) that is much closer to the ideal values. This method can be used on top of other noise mitigation techniques \cite{cai2023quantum} as well in order to reduce errors that bring the system out of the charge-singlet subspace. However, CSM cannot correct for errors that occur within the singlet subspace.    

\begin{figure}[ht]
    \centering
    \includegraphics[width=1.0\linewidth]{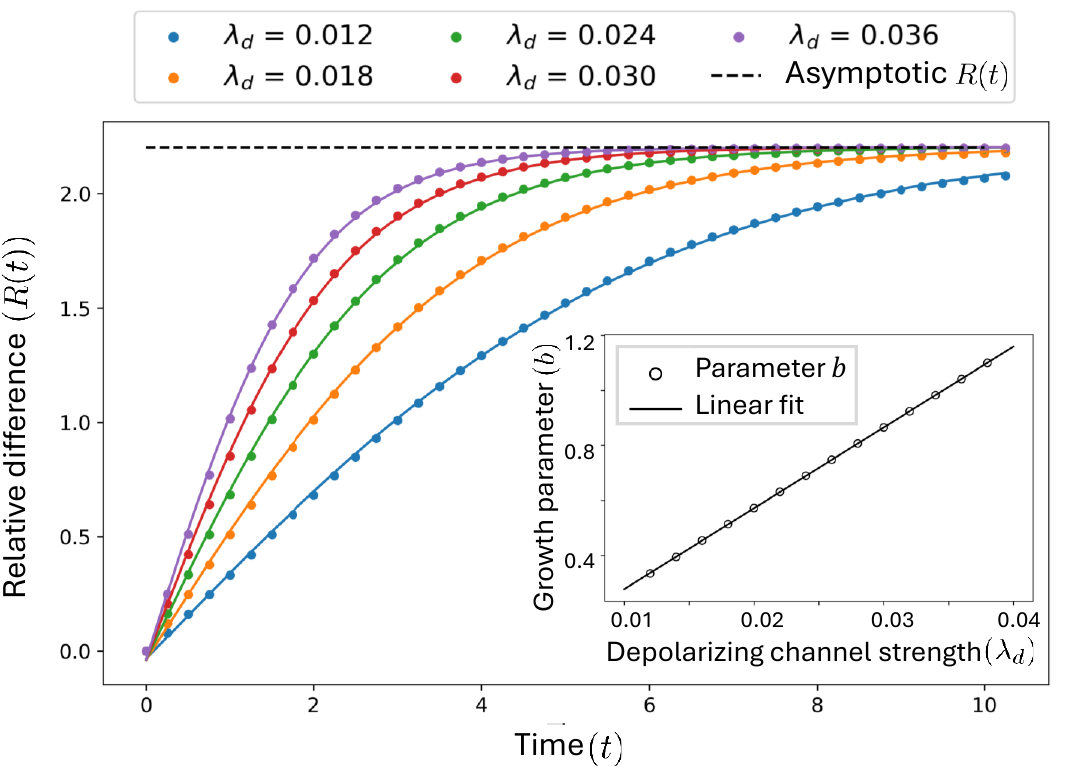}
    \caption{\textbf{Relation between noise strength and performance of the CSM technique for time evolution of a SU(2) unit cell.} The behavior of the quantity $R(t)$ with time is shown here for different depolarizing noise strength $\lambda_d$ for a simulation performed in the same setting as described in Sec~\ref{sec:noisy-time-evolve}. The quantity $R(t)$ approaches an asymptotic value (dashed line) as the density matrix gradually approaches the maximally mixed state. The data points are fitted with a sigmoid function $f(x)=a/(1+e^{-bx})+c$, where the parameter $b$ determines how fast the density matrix approaches the maximally mixed state. The fitted sigmoid functions are shown as solid curves. The inset shows the linear dependence of the fit parameter $b$ on $\lambda_d$.
}
    \label{fig:sigmoid_fit}
\end{figure}

In addition to mitigating noise to improve the accuracy of observable expectation values, the CSM approach also enables the extraction of the two-qubit depolarizing noise channel strength as a byproduct of the time-evolution simulation. For the noisy simulation of the time evolution of a SU(2) unit cell, this can be illustrated by considering the behavior of the quantity
\begin{align}
   R(t) &=  \frac{{\rm Tr}(\hat{\rho}_0(t)\ket{\psi_t}\bra{\psi_t}) - {\rm Tr}(\hat{\rho}(t)\ket{\psi_t}\bra{\psi_t})}{{\rm Tr}(\hat{\rho}(t)\ket{\psi_t}\bra{\psi_t})}\nonumber\\
   &= \frac{1}{{\rm Tr}(\hat{\rho}(t)\hat{K})} - 1\;, \label{eq:rel-error-K}
\end{align}
with time. $R(t)$ is the relative difference of fidelity between the prepared state $\hat{\rho}(t)$ and the projected density matrix $\hat{\rho}_0(t)$ with the target state $\ket{\psi_t}$.
Equation~(\ref{eq:rel-error-K}) can be evaluated solely from diagonal measurements of the noisy density matrix and, for a fixed initial state, depends at any given time only on the strength of the noise channel. As time increases, $R(t)$ approaches a saturation value as the density matrix gradually evolves towards the maximally mixed state in the presence of noise
\begin{equation}
    R(t\rightarrow\infty) = \frac{2^{N_q}}{{\rm Tr}(\hat{K})} - 1\;.
\end{equation}
For a SU(2) unit cell, ${\rm Tr}(\hat{K}) = 5$ and $R(t\rightarrow\infty) = 2.2$. The rate of growth of the quantity $R(t)$ to reach this saturation value depends on the strength ($\lambda_d$) of the noise channel. Fig.~\ref{fig:sigmoid_fit} shows the behavior of $R(t)$ with time for different $\lambda_d$. The growth rate can be determined from the sigmoid form of the curve. The data are fit to the sigmoid function $f(x) = a/(1+e^{-bx})+c$, and , and the resulting growth parameter $b$ is found to increase linearly with the channel strength $\lambda_d$. The linear relationship shown in the inset of Fig.~\ref{fig:sigmoid_fit} enables estimation of the depolarizing noise strength \( \lambda_d \) based on the experimentally determined value of \( b \).

\subsection{Ground state preparation}\label{sec:ground-state-prep}
\begin{figure}[h]
    \centering
    \includegraphics[width=1.0\linewidth]{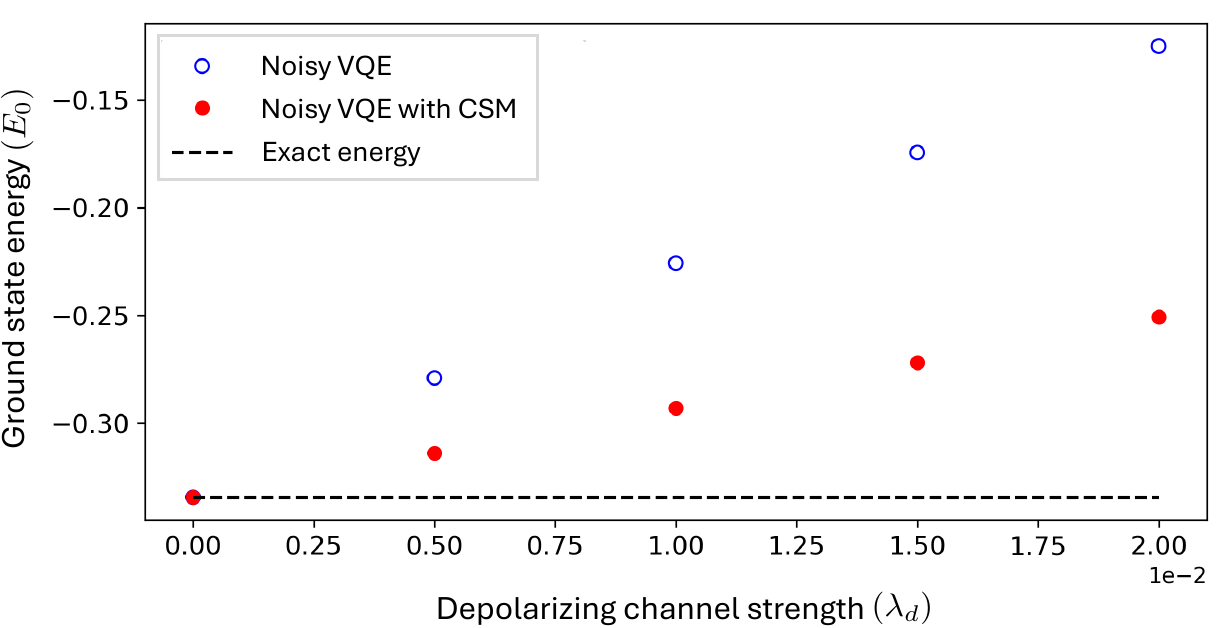}
    \caption{\textbf{Ground state energy determination using CSM for a SU(2) unit cell with matter.} As the depolarizing noise strength $\lambda_d$ associated with each two-body gate is increased, the ground state prepared by the variational protocol has decreasing fidelity with the true ground state, leading to larger error. Some errors can be compensated for by the CSM protocol, which gives us a better estimate of the ground state energy.}
    \label{fig:gs_meson_su2}
\end{figure}

A closely related application of using charge-singlet measurements for mitigating non-charge-singlet errors induced from noisy gates is the preparation of ground states. Like the previous example, a symmetry preserving circuit can also prepare non-charge-singlet states due to errors. Using the CSM technique one can obtain better estimates of ground state properties by projecting out the non-charge-singlet error. Here we show an example for the ground state search of a SU(2) unit cell in (1+1)-D in the presence of matter. We have used the variational circuit (before any reduction) provided in Ref.~\cite{atas2021} (see Appendix~\ref{app:circuits}). In the absence of noise, the circuit is able to prepare the charge-singlet ground state with a high degree of accuracy, as is evident from the exact match of the VQE estimation with the actual ground state energy obtained from exact diagonalization (Fig.~\ref{fig:gs_meson_su2}). 

We now apply the same noise model as in the case of time evolution, i.e., each two-qubit gate is followed by a two-qubit depolarizing channel. As the strength of the depolarizing noise channel increases, the prepared state becomes increasingly mixed, resulting in a deviation from the true ground state energy. By using charge-singlet measurements, we can determine the ground state energy with a higher accuracy (see Fig.~\ref{fig:gs_meson_su2}).  

\section{Dimension of singlet subspace}\label{sec:dimension_subspace}
In this section, we show how to compute the size of the relevant singlet subspace for SU(2) and SU(3) lattice gauge theories in (1+1)-D using the analytic expression for $\hat{K}$ derived in Sec.~\ref{sec:proj_operator-K}. We recall that the trace of the idempotent projection operator $\hat{P}_0$ in Eq.~(\ref{eq:P_0-general-finite}) provides the dimension of the singlet subspace, which corresponds to the multiplicity of the eigenvalue 1. 
Since $\hat{P}_0$ can be replaced by the diagonal operator $\hat{K}$ inside the trace, $\operatorname{Tr}(\hat{K})$ yields the dimension $\mathcal{D}$ of the singlet subspace.
\\



\begin{figure}[h]
    \centering
    \includegraphics[width=1.0\linewidth]{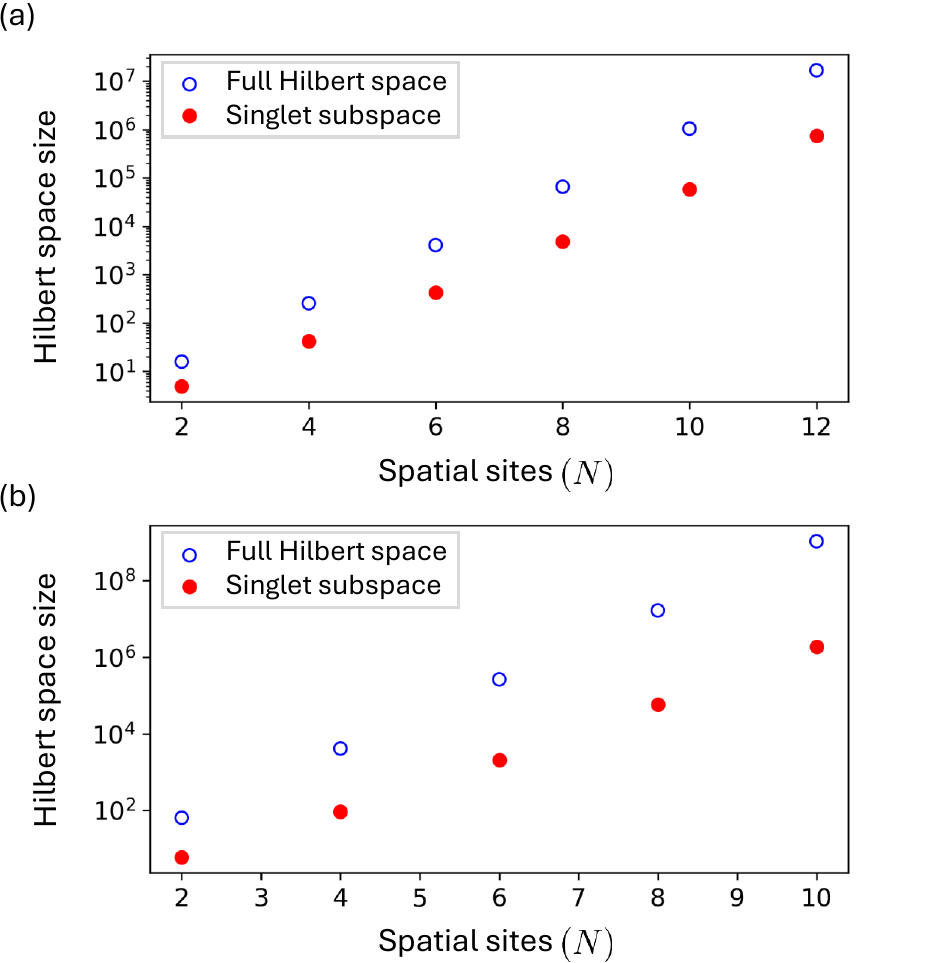}
    \caption{\textbf{Singlet subspace dimension for SU(2) and SU(3) LGT in (1+1)-D.} Comparison of the dimension of the singlet subspace with the full Hilbert space for (a) SU(2) and (b) SU(3) LGT in (1+1)-D. The $y$-axis is in log scale showing that the dimension of the singlet subspace also grows exponentially with system size.}
    \label{fig:dimension-singlet}
\end{figure}

\noindent\textit{SU$(2)$ LGT ---} For SU$(2)$ LGT in (1+1)-D, the diagonal projection operator can be written as
\begin{equation}
    \hat{K} = \frac{1}{2\pi}\int_0^{4\pi}  d\alpha \, \sin^2(\alpha/2)\, e^{i\alpha\, \hat{Q}_{tot}^z}\;.
\end{equation}
Using the explicit form of the diagonal charge operator for $N$ spatial sites given in Eq.~(\ref{eq:Qz_tot}), the trace is expressed as
\begin{align}
    \mathcal{D} = {\rm Tr}(\hat{K}) &= \frac{4^N}{2\pi}\int_0^{4\pi}  d\alpha \, \sin^2(\alpha/2)\, \cos^{2N}(\alpha/4) \nonumber\\
    & = \frac{(2N+2)!}{(N+1)!\, (N+2)!} \;.\label{eq:su2-dimension-physical}
\end{align}
The expression above gives the dimension of the charge-singlet Hilbert space, whereas the dimension of the full Hilbert space is given by $2^{2N}$ (Fig.~\ref{fig:dimension-singlet}a).\\

\noindent\textit{SU$(3)$ LGT ---} In the case SU$(3)$ LGT in (1+1)-D, there are two conserved diagonal charges $\hat{Q}_{tot}^3$ and $\hat{Q}_{tot}^8$, which spans the Cartan subalgebra. The diagonal SU$(3)$ projection operator takes the form
\begin{equation}
    \hat{K} = \int d\mu(a,b)\, e^{ia\hat{Q}_{tot}^3 + 2ib\hat{Q}_{tot}^8/\sqrt{3}}\;.
\end{equation}
Here, $d\mu(a,b)$ denotes the invariant measure of the Cartan subgroup of SU(3), as given in Eq.~(\ref{eq:SU3-parametrization}). 
To find the trace of the projection operator we note that
\begin{align}
    {\rm Tr}(e^{ia\hat{Q}_{tot}^3 + 2ib\hat{Q}_{tot}^8/\sqrt{3}}) =  & 4^N\cos^N(b/3) \times\nonumber\\
    &\left[\cos(a/2) + \cos(b/3) \right]^N \;.
\end{align}
The dimension of the singlet Hilbert space for a given \( N \) can then be computed through numerical integration. For \( N = 2, 4, 6 \), the dimension \( \mathcal{D} = 6, 92, 2074 \), respectively, illustrating the rapid growth of the singlet subspace with system size (see Fig.~\ref{fig:dimension-singlet}b). In comparison, the dimension of the full Hilbert space scales as \( 2^{3N} \).

Although we focus on calculating the singlet Hilbert space dimension for SU(2) and SU(3) lattice gauge theories in (1+1)-D using the trace of the projection operator \( \hat{K} \), the relation \( \mathcal{D} = \operatorname{Tr}(\hat{P}_0) \) holds generally for all discrete and continuous gauge groups.




\section{Conclusions}
In this article, we presented an alternative approach to implementing color-neutrality constraints in non-Abelian lattice gauge theories. Rather than enforcing these constraints during state preparation, our method uses a projection operator that applies color-neutrality at the stage of observable measurement. The resulting charge-singlet measurement protocol proves to be a versatile tool—not only for satisfying gauge constraints but also for mitigating noise in symmetry-preserving circuits, including circuits for trotterized time evolution.

A key advantage of this approach is that, since the projection is applied during postprocessing, it allows the use of highly expressive variational ansatzes for eigenstate and thermal state preparation. These ansatzes can be either classical, such as tensor networks (TN) and neural networks, or quantum, such as variational quantum eigensolvers (VQE). Implementing gauge constraints in the architecture of a tensor network or a neural network is an active area of research and the CSM approach provides an alternative to these method, where the architecture can remain arbitrary. 

The CSM approach can be realized across various quantum platforms, as it requires only single-qubit rotation gates. The method was experimentally implemented in Ref.~\cite{than2024phase}, where it was used to map the phase diagram of (1+1)-D QCD using a trapped-ion device. This technique is also suitable for systems with individual qubit control, potentially enabling the simulation of higher-dimensional lattice gauge theories without the need for directly implementing non-Abelian gauge constraints.

Additionally, we explored how the diagonal projection operator connects with thermodynamic quantities, such as entropy. As the relationship between lattice gauge theories and thermodynamics gains increasing attention \cite{davoudi2024quantum,majidy2023noncommuting}, experimental verification of theoretical predictions requires measuring quantities like internal energy, entropy, and work in the charge-singlet subspace. Our method has the potential to offer a practical route for such measurements on quantum devices.

In summary, the charge-singlet measurement toolbox we propose opens new avenues for exploring eigenstate and thermal state properties in lattice gauge theories. It also provides a means of noise mitigation in quantum circuits and facilitates exploration of the intersection between quantum thermodynamics and gauge theory—while being readily implementable in both classical and quantum simulations.


\section*{Acknowledgments}
We thank Yasar Atas for his important theory contributions to developing the charge-singlet measurement framework introduced in \cite{than2024phase}. We are also grateful to Jinglei Zhang and to our experimental collaborators—especially Anton Than, Alaina Green, and Norbert Linke—for demonstrating these ideas using trapped ions and for their invaluable conceptual input toward making charge-singlet measurements practical. This research was supported by the Natural Sciences and Engineering Research Council of Canada (NSERC), the Canada First Research Excellence Fund (CFREF, Transformative Quantum Technologies), New Frontiers in Research Fund (NFRF), Ontario Early Researcher Award, and the Canadian Institute for Advanced Research (CIFAR).

\appendix

\renewcommand\thefigure{A.\arabic{figure}}    
\setcounter{figure}{0} 
\section{Hamiltonians for SU(2) and SU(3) LGTs in (1+1)-D} \label{app:unit-cell-hamiltonian}
In this appendix, we provide the form of the SU(2) and SU(3) Hamiltonian in (1+1)-D with open boundary conditions mapped onto qubits using Jordan-Wigner transformation. Details of the derivation are given in \cite{atas2021,atas2023}. Furthermore, we provide the expression for all the conserved charges that were used in the main text.

\subsection{Hamiltonian of SU(2) LGT in (1+1)-D}\label{appsec:su2-Ham}
The qubit Hamiltonian for SU(2) LGT in the staggered formulation consists of the following contributions as mentioned in the main text Eq.~(\ref{eq:generic-Hamiltonian})
\begin{equation}
	\hat{H} = \hat{H}_{kin}+m\hat{H}_{m} + g^2\hat{H}_{el} - \mu \hat{H}_{chem}, \label{appendix:hamiltonian_general}
\end{equation}
where $\hat{H}_{kin}$ is the kinetic energy, $\hat{H}_{m}$ is the mass term, $\hat{H}_{el}$ is the color electric energy and $\hat{H}_{chem}$ is the chemical potential energy. Here $m$, $\mu$ and $g^2$ are the dimensionless mass, chemical potential and coupling constant, respectively. After the elimination of the gauge fields and mapping to qubits, the terms take the following forms
{\allowdisplaybreaks
\begin{align}
	\hat{H}_{kin} &= -\frac{1}{2}\sum_{n=1}^{N-1}\left( \hat{\sigma}_{2n-1}^{+}\hat{\sigma}_{2n}^{z}\hat{\sigma}_{2n+1}^{-} + \hat{\sigma}_{2n}^{+}\hat{\sigma}_{2n+1}^{z} \hat{\sigma}_{2n+2}^{-} + \text{H.c.}\right),  \label{eq:qubitkinetic}\\
    \hat{H}_{m} &= \sum_{n = 1}^{N} \left(\frac{(-1)^{n}}{2}\left( \hat{\sigma}_{2n-1}^{z}  + \hat{\sigma}_{2n}^{z}\right)+1 \right),\label{eq:qubitmass}\\
\notag	\hat{H}_{el}&=\frac{3}{16}\sum_{n=1}^{N-1}(N-n)(1-\hat{\sigma}_{2n-1}^{z}\hat{\sigma}_{2n}^{z})\\  \notag &+\frac{1}{16}\sum_{n=1}^{N-2}\sum_{m>n}^{N-1}(N-m)\left(\hat{\sigma}_{2n-1}^{z}-\hat{\sigma}_{2n}^{z}\right)\left(\hat{\sigma}_{2m-1}^{z}-\hat{\sigma}_{2m}^{z}\right) \\
&+\frac{1}{2}\sum_{n=1}^{N-2}\sum_{m>n}^{N-1}(N-m)\left(\hat{\sigma}_{2n-1}^{+}\hat{\sigma}_{2n}^{-}\hat{\sigma}_{2m}^{+}\hat{\sigma}_{2m-1}^{-}+\mathrm{H.c.}\right),\label{eq:qubitelectric} \\
\hat{H}_{chem}&=\frac{1}{4}\sum_{n=1}^{2N}\hat{\sigma}_{n}^{z},\label{eq:qubitchem}
\end{align}}
where $\hat{\sigma}_{n}^{x,y,z}$ are the usual Pauli matrices, $\hat{\sigma}_{n}^{\pm}=(\hat{\sigma}_{n}^{x}\pm i \hat{\sigma}_{n}^{y})/2$, and $N$ is the number of lattice sites. Due to the presence of two color components of the fermionic fields, we need $2N$ qubits for the simulation of the Hamiltonian in Eq.~(\ref{appendix:hamiltonian_general}). For all the SU(2) simulations done in this article, we have used the parameter values $m = 0.5, g^2=0.5$. For simulations where the chemical potential $\mu$ is not varied,  i.e., for all results in Sec.~\ref{sec:noise-mitigation}, it is taken to be zero. 

The conserved charges take the following expressions in terms of Pauli operators  \cite{atas2021}
\begin{align}
\hat{Q}_{tot}^{x}&=\frac{1}{2}\sum_{n=1}^{N}\left( \hat{\sigma}_{2n-1}^{+}\hat{\sigma}_{2n}^{-}+\mathrm{H.c.}\right), \label{appendix:su2_charges_x}\\
\hat{Q}_{tot}^{y}&=\frac{i}{2}\sum_{n=1}^{N}\left( \hat{\sigma}_{2n-1}^{-}\hat{\sigma}_{2n}^{+}-\mathrm{H.c.}\right), \label{appendix:su2_charges_y}\\
\hat{Q}_{tot}^{z}&=\frac{1}{4}\sum_{n=1}^{N}\left( \hat{\sigma}_{2n-1}^{z}-\hat{\sigma}_{2n}^{z}\right). \label{appendix:su2_charges_z}
\end{align}

\subsection{Hamiltonian of SU(3) LGT in (1+1)-D}
For SU(3), the Hamiltonian is of the same form as in Eq.~(\ref{appendix:hamiltonian_general}), with each term given by the following
{\allowdisplaybreaks
\begin{align}
    \hat{H}_{kin}=\notag \frac{1}{2}\sum_{n=1}^{N-1}(-1)^n& \left(\hat{\sigma}_{3n-2}^{+}\hat{\sigma}_{3n-1}^{z}\hat{\sigma}_{3n}^{z} \hat{\sigma}_{3n+1}^{-}\right. \\
   &\notag -\left. \hat{\sigma}_{3n-1}^{+}\hat{\sigma}_{3n}^{z}\hat{\sigma}_{3n+1}^{z} \hat{\sigma}_{3n+2}^{-}\right. \\
   & +\left. \hat{\sigma}_{3n}^{+}\hat{\sigma}_{3n+1}^{z}\hat{\sigma}_{3n+2}^{z} \hat{\sigma}_{3n+3}^{-} +\operatorname{H.c.}\right), \label{kinetic_ham_qubit}
\end{align}
\begin{align}
\hat{H}_{m}=&\frac{1}{2}\sum_{n=1}^{N}\left[(-1)^{n}\left(\hat{\sigma}_{3n-2}^{z}+\hat{\sigma}_{3n-1}^{z}+\hat{\sigma}_{3n}^{z}\right)+3\right]. \label{mass_ham_qubit}\\
    \hat{H}_{el}=&\notag \frac{1}{3}\sum_{n=1}^{N-1}(N-n) \\
   & \notag\times  
     \left( 3-\hat{\sigma}_{3n-2}^{z}\hat{\sigma}_{3n-1}^{z}-\hat{\sigma}_{3n-2}^{z}\hat{\sigma}_{3n}^{z}-\hat{\sigma}_{3n-1}^{z}\hat{\sigma}_{3n}^{z}\right)\\\notag &+\sum_{n=1}^{N-2}\sum_{m=n+1}^{N-1}\left[ (N-m)\left( \hat{\sigma}_{3n-2}^{+}\hat{\sigma}_{3n-1}^{-}\hat{\sigma}_{3m-1}^{+}\hat{\sigma}_{3m-2}^{-}\right. \right. \\
   \notag & +\left. \hat{\sigma}_{3n-1}^{+}\hat{\sigma}_{3n}^{-}\hat{\sigma}_{3m-1}^{-}\hat{\sigma}_{3m}^{+}+\operatorname{H.c.}\right)(-1)^{n+m} \\
   \notag & +(N-m)\left(\hat{\sigma}_{3n-2}^{+}\hat{\sigma}_{3n-1}^{z}\hat{\sigma}_{3n}^{-}\hat{\sigma}_{3m-2}^{-}\hat{\sigma}_{3m-1}^{z}\hat{\sigma}_{3m}^{+} +\operatorname{H.c.}\right) \\
   \notag &-\frac{1}{12}(N-m) \hat{\sigma}_{3m-2}^{z}(\hat{\sigma}_{3n-1}^{z}+\hat{\sigma}_{3n}^{z}-2\hat{\sigma}_{3n-2}^{z}) \\
   \notag &-\frac{1}{12}(N-m) \hat{\sigma}_{3m-1}^{z}(\hat{\sigma}_{3n}^{z}+\hat{\sigma}_{3n-2}^{z}-2\hat{\sigma}_{3n-1}^{z})\\
    & \left. -\frac{1}{12}(N-m) \hat{\sigma}_{3m}^{z}(\hat{\sigma}_{3n-2}^{z}+\hat{\sigma}_{3n-1}^{z}-2\hat{\sigma}_{3n}^{z}) \right], \label{qubit_electric_general}
\end{align}
\begin{equation}
    \hat{H}_{chem}=\frac{1}{6}\sum_{n=1}^{3N}\hat{\sigma}_{n}^{z}. \label{qubit_baryon_number}
\end{equation}}
The same parameters $m=g^2=0.5$ are used to define the Hamiltonian. Since there are three color components for the fermionic fields, we need $3N$ qubits to simulate it where $N$ is the number of staggered spatial sites.

There are eight conserved charges for SU(3) LGT, out of which two are diagonal and the rest are non-diagonal. The diagonal form of the two charges $\hat{Q}_{n}^{3}$ and $\hat{Q}_{n}^{8}$ are evident from the presence of only $\hat{\sigma}_{n}^{z}$ operators in their expressions as given below.
\allowdisplaybreaks
\begin{align}
    \hat{Q}_{n}^{1}&=\frac{(-1)^{n}}{2}\left( \hat{\sigma}_{3n-2}^{+}\hat{\sigma}_{3n-1}^{-}+\operatorname{H.c.}\right)\label{qubit_nonabeliancharge1}, \\
    \hat{Q}_{n}^{2}&=\frac{i(-1)^{n}}{2}\left( \hat{\sigma}_{3n-1}^{+}\hat{\sigma}_{3n-2}^{-}-\operatorname{H.c.}\right)\label{qubit_nonabeliancharge2},\\
    \hat{Q}_{n}^{3}&=\frac{1}{4}\left( \hat{\sigma}_{3n-2}^{z}-\hat{\sigma}_{3n-1}^{z}\right)\label{qubit_nonabeliancharge3},\\
    \hat{Q}_{n}^{4}&=-\frac{1}{2}\left( \hat{\sigma}_{3n-2}^{+}\hat{\sigma}_{3n-1}^{z}\hat{\sigma}_{3n}^{-}+\operatorname{H.c.}\right)\label{qubit_nonabeliancharge4}, \\
    \hat{Q}_{n}^{5}&=\frac{i}{2}\left( \hat{\sigma}_{3n-2}^{+}\hat{\sigma}_{3n-1}^{z}\hat{\sigma}_{3n}^{-}-\operatorname{H.c.}\right)\label{qubit_nonabeliancharge5}, \\
    \hat{Q}_{n}^{6}&=\frac{(-1)^n}{2}\left( \hat{\sigma}_{3n-1}^{+}\hat{\sigma}_{3n}^{-}+\operatorname{H.c.}\right)\label{qubit_nonabeliancharge6},\\
    \hat{Q}_{n}^{7}&=\frac{i(-1)^n}{2}\left( \hat{\sigma}_{3n}^{+}\hat{\sigma}_{3n-1}^{-}-\operatorname{H.c.}\right)\label{qubit_nonabeliancharge7},\\
    \hat{Q}_{n}^{8}&=\frac{1}{4\sqrt{3}}\left( \hat{\sigma}_{3n-2}^{z}+\hat{\sigma}_{3n-1}^{z}-2\hat{\sigma}_{3n}^{z}\right)\label{qubit_nonabeliancharge8}.
\end{align}

\section{Singlet states for SU(2) and SU(3) unit cells in (1+1)-D}\label{appsec:singlet-states}
In this section, we briefly describe how the single unit cell for SU(2) and SU(3) LGT is represented by qubits. We then provide the qubit form of the singlet basis states in the strong-coupling regime.

\begin{figure}[h]
    \centering
    \includegraphics[width=0.5\linewidth]{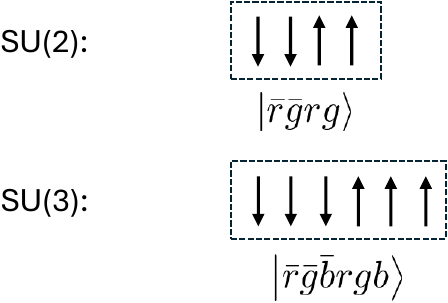}
    \caption{\textbf{Unit cell occupancy of SU(2) and SU(3) LGT in (1+1)-D.} Mapping of a fully occupied unit cell to qubits are shown for SU(2) and SU(3) unit cell $(N=2)$. The first half of the cell accommodates antiquarks and the other half is occupied by quarks. A vacant site is represented by the opposite spin of the occupied site. For example, a single red quark in a SU(3) unit cell is denoted by $\left|\uparrow \uparrow\uparrow\uparrow\downarrow\downarrow\right\rangle$.}
    \label{fig:unit-cell-occupancy}
\end{figure}

For SU(2) LGT, a single unit cell is mapped on to four qubits: the first two qubits $(N=1)$ representing the anti-quark sites and the last two qubits denoting the quark sites. The occupation of each site is defined in a way such that it respects the Jordan-Wigner transformation that mapped the fermions to qubits (Fig.~\ref{fig:unit-cell-occupancy}). For SU(3) LGT, the unit cell is represented by six qubits, with the first three qubits reserved for the anti-quark and the last three qubits representing quarks (Fig.~\ref{fig:unit-cell-occupancy}). 

In Sec.~\ref{sec:decomp_irreducible}, we discussed the strong-coupling singlet basis states that can be accommodated in a single unit cell for SU(2) and SU(3) LGT. The number of such basis states determine the dimension of the singlet subspace. As mentioned in  Sec.~\ref{sec:decomp_irreducible}, the strong-coupling singlet states for SU(2) are the vacuum $(|vac\rangle)$, baryon $(|B\rangle)$, antibaryon $(|\bar{B}\rangle)$, meson $(|M\rangle)$, and baryonium $(|\bar{B}B\rangle)$. For SU(3), there is one more singlet state, which is called a tetraquark $(|T\rangle)$state. We show below the singlet states in terms of the qubit representation.
\begin{align*}
 {\rm SU}(2) : \begin{cases}
      & |vac\rangle = |0011\rangle \\
    & |B\rangle = |0000\rangle \\
    & |\bar{B}\rangle = |1111\rangle\\
    & |M\rangle = \frac{1}{\sqrt{2}}\left(|1001\rangle + |0110\rangle \right)\\
    & |\bar{B}B\rangle = |1100\rangle
 \end{cases}  \;,
\end{align*}

\begin{align*}
 {\rm SU}(3) : \begin{cases}
      & |vac\rangle = |000111\rangle \\
    & |B\rangle = |000000\rangle \\
    & |\bar{B}\rangle = |111111\rangle\\
    & |M\rangle = \frac{1}{\sqrt{3}}\left(|100011\rangle + |010101\rangle + |001110\rangle\right)\\
    & |T\rangle = \frac{1}{\sqrt{3}} \left(|110001\rangle + |011100\rangle + |101010\rangle \right) \\
    & |\bar{B}B\rangle = |111000\rangle
 \end{cases}  \;.
\end{align*}
Here, we have used the notation $|\hspace{-0.2em}\uparrow\rangle \equiv |0\rangle$ and $|\hspace{-0.2em}\downarrow\rangle \equiv |1\rangle$.

\section{Variational quantum thermalizer ansatz for thermal state simulations}\label{appsec:VQT-circuit}
Here we provide a brief description of the circuit used to prepare thermal states for a SU(2) unit cell following \cite{than2024phase}. Details about the circuit and optimization process is provided in \cite{than2024phase}. The result of the optimization performed on this circuit is used for determining the electric field energy in Sec.~\ref{sec:thermal-states} and the entropy of the singlet density matrix in Sec.~\ref{sec:GI-entropy}. 

\begin{figure}[h]
    \centering
    \includegraphics[width=1.0\linewidth]{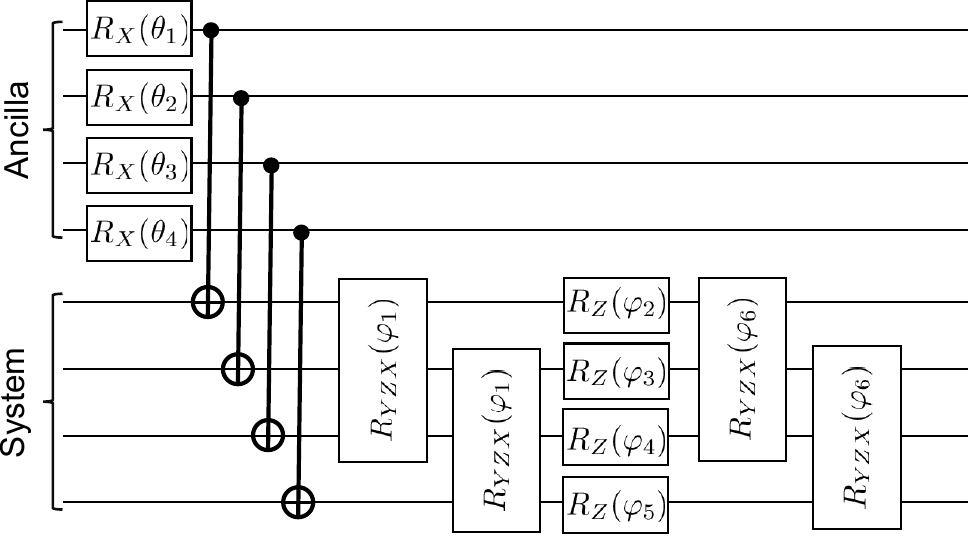}
    \caption{\textbf{Variational quantum thermalizer (VQT) ansatz for SU(2) unit cell.} The circuit is divided into two parts: an ancilla register which prepares the probability distribution and a system register which creates the eigenstates of the density matrix. There are ten variational parameters which are optimized to find the minimum free energy state. In the circuit diagram $R_X(\theta_i) = e^{i\theta_i\hat{\sigma}_x}/2$, $R_Z(\varphi_i) = e^{i\varphi_i\hat{\sigma}_z}/2$ are single qubit $x$ and $z$ rotations. $R_{YZX}(\varphi_i) = e^{i\varphi_i\hat{\sigma}_y\otimes \hat{\sigma}_z\otimes\hat{\sigma}_x}/2$ is a parametrized three-qubit gate that can be decomposed into single and two-qubit gates.}
    \label{fig:su2-thermal}
\end{figure}

Fig.~\ref{fig:su2-thermal} shows the circuit used to find the density matrix in the reducible representation that minimizes the free energy. The energy is determined by measuring the Hamiltonian $\hat{H}$ on the system register. The entropy $S$ can be found by using the analytical expression
\begin{align}
\notag S(\boldsymbol{\theta})=-\sum_{i} & \left[\cos^2{(\theta_i/2)}\log(\cos^2{(\theta_i/2)}) \right. \\ &+\left.\sin^2{(\theta_i/2)}\log(\sin^2{(\theta_i/2)})\right] \;.\label{eq:entropy_analytical}
\end{align}
Here, the single-qubit rotation gate angles in the ancilla register can be used to determine the entropy. An alternative way would be to measure the ancilla register and evaluate $S=-\sum_np_n\log p_n$ where $p_n$ are the probabilities of getting the computational basis state $\ket{n}$.

The minimization of the free energy creates an optimized circuit with optimal parameters $\left(\boldsymbol{\theta^*},\boldsymbol{\varphi^*}\right)$. The optimized circuit creates a Gibbs state $\hat{\rho}$, which is the reducible density matrix that we use for performing the charge-singlet measurements. 

\section{Alternative derivation of Eq.~(\ref{eq:exp-val-K})}\label{app:alternative-deriv}
In this appendix, we provide an alternative derivation of Eq.~(\ref{eq:exp-val-K}) following Refs.~\cite{greiner2012quantum,elze1986quantum}. Let us consider an observable $\hat{\Omega}$ that needs to be measured on the charge-singlet subspace. Since $\hat{\Omega}$ is an observable, the operator should commute with all elements of the gauge group. We now define the following expression
\begin{equation}
    \langle\tilde{\Omega}\rangle  = {\rm Tr}\left[\hat{\Omega} e^{-i\alpha_k\hat{Q}^k}\right]\;,
\end{equation}
where the sum over $k$ is implied in the exponent (Einstein's convention) and $\hat{Q}^k$ are the diagonal conserved charges belonging to the Cartan subalgebra of the Lie group. If the operator $\hat{\Omega}$ is expressed in a reducible representation, we can write the trace as 
\begin{equation}
    \langle\tilde{\Omega}\rangle = \sum_{p,m} \bra{p,m} \hat{\Omega} e^{i\alpha_k\hat{Q}^k} \ket{p,m} \;.
\end{equation}
Here, $p$ denotes an irreducible representation and $m$ are the quantum numbers associated with the diagonal charges, e.g., for SU(3), $m$ are the quantum numbers associated with the eigenvalues of $\hat{Q}^3$ and $\hat{Q}^8$ for a given representation $(p,q)$.

We now introduce the decomposition of identity in terms of a complete set of states to get 
\begin{equation}
    \langle\tilde{\Omega}\rangle = \sum_{p,m} \sum_{p',m'} \bra{p,m} \hat{\Omega}  \ket{p',m'} \bra{p',m'}e^{i\alpha_k\hat{Q}^k}\ket{p,m}\;. \label{eq:tilde-omega-completeness}
\end{equation}

Since $e^{-i\alpha_k\hat{Q}^k}$ does not change the representation, nor the quantum numbers $m$, we have $\bra{p',m'}e^{-i\alpha_k\hat{Q}^k}\ket{p,m} = \bra{p,m}e^{-i\alpha_k\hat{Q}^k}\ket{p,m}\, \delta_{p,p'}\delta_{m,m'}$. We now note that the definition of a character of a particular irreducible representation $p$
\begin{equation}
    \chi^p(\boldsymbol{\alpha}) = \sum_m \bra{p,m}e^{i\alpha_k\hat{Q}^k}\ket{p,m}
\end{equation}
where the elements of the vector $\boldsymbol{\alpha}$ are $\alpha_k$. Since $\hat{\Omega}$ does not depend on $m$ (as it preserves the color symmetry), we can then write Eq.~(\ref{eq:tilde-omega-completeness}) as
\begin{equation}
    \langle\tilde{\Omega}\rangle = \sum_{p} \bra{p} \hat{\Omega}  \ket{p} \,\chi^p(\boldsymbol{\alpha})
\end{equation}
The first factor in the equation above is related to the trace of the observable over a particular irreducible representation, i.e., $\bra{p} \hat{\Omega}  \ket{p} = {\rm Tr}_p(\hat{\Omega})/{\rm dim}(p)$. The dimensionality of the representation $p$ appears in the denominator as ${\rm Tr}_p(\hat{\Omega}) = \sum_{m_p=1}^{{\rm dim}(p)} \bra{p,m_p}\hat{\Omega}\ket{p,m_p} = \sum_{m_p=1}^{{\rm dim}(p)} \bra{p}\hat{\Omega}\ket{p}$, where we have used the fact that $\hat{\Omega}$ does not change $m_p$. We can then write Eq.~(\ref{eq:tilde-omega-completeness}) in a more simplified form
\begin{equation}
    \langle\tilde{\Omega}\rangle = \sum_p \frac{{\rm Tr}_p(\hat{\Omega})}{{\rm dim}(p)} \,\chi^p(\boldsymbol{\alpha})\;.
\end{equation}
Using the orthogonality relation of characters, we can then extract ${\rm Tr}_p(\hat{\Omega})$,
\begin{equation}
    {\rm Tr}_p(\hat{\Omega}) = {\rm dim}(p) \int d\mu(\boldsymbol{\alpha})\, \langle\tilde{\Omega}\rangle\, \chi^p(\boldsymbol{\alpha}) \label{eq:generic-p-omega-expectation}
\end{equation}

In Eq.~(\ref{eq:generic-p-omega-expectation}), we want to only find the expectation value on the charge-singlet subspace, i.e., for $p=0$. For the singlet subspace, ${\rm dim}(p) = 1$, and $\chi^0(\boldsymbol{\alpha)}=1$, which yields
\begin{equation}
    {\rm Tr}_0(\hat{\Omega}) =  \int d\mu(\boldsymbol{\alpha})\, \langle\tilde{\Omega}\rangle\, \,. \label{eq:singlet-omega-expectation}
\end{equation}
To retrieve Eq.~(\ref{eq:exp-val-K}) from Eq.~(\ref{eq:singlet-omega-expectation}), we take $\hat{\Omega} = e^{-\beta \hat{H}}\hat{O}$ to get
\begin{equation}
    {\rm Tr}_0(e^{-\beta \hat{H}}\hat{O}) =  \int d\mu(\boldsymbol{\alpha})\, {\rm Tr}\left[e^{-\beta \hat{H}}\hat{O}\,e^{-i\alpha_k\hat{Q}^k} \right]\, \label{eq:intermediate-tr-relation}
\end{equation}
For a given observable $\hat{O}$, the expectation value (on the singlet subspace) at a finite temperature is defined as $\langle\hat{O\rangle}_{0} = {\rm Tr}_0(e^{-\beta \hat{H}}\hat{O})/Z_0$, where $Z_0$ is the color-singlet partition function $Z_0 = {\rm Tr}_0 (e^{-\beta \hat{H}})$. Using Eq.~(\ref{eq:intermediate-tr-relation}) we can then write
\begin{align}
    \langle\hat{O\rangle}_{0} &= \frac{Z}{Z_0} \int d\mu(\boldsymbol{\alpha})\, {\rm Tr}\left[Z^{-1}e^{-\beta \hat{H}}\hat{O}\,e^{-i\alpha_k\hat{Q}^k} \right] \nonumber\\
    &= \frac{Z}{Z_0} {\rm Tr}\left[\hat{\rho}\hat{O}\hat{K}\right]
\end{align}
where we have used the definition of $\hat{K} = \int d\mu(\boldsymbol{\alpha})\,e^{-i\alpha_k\hat{Q}^k}$ and the reducible density matrix $\hat{\rho} = Z^{-1}e^{-\beta \hat{H}}$. $Z$ is the partition function corresponding to the reducible representation. We can use Eq.~(\ref{eq:z0-z-K-relation}) to replace the prefactor $Z/Z_0$ to retrieve Eq.~(\ref{eq:exp-val-K})
\begin{equation}
    \langle\hat{O\rangle}_{0} = \frac{{\rm Tr}(\hat{\rho}\hat{O}\hat{K})}{{\rm Tr}(\hat{\rho}\hat{K})}\;.
\end{equation}

\section{Time evolution and ground state preparation circuits} \label{app:circuits}
In this appendix, we provide more details on the circuits used for time evolution of a SU(2) unit cell in Sec.~\ref{sec:noisy-time-evolve} and for ground state preparation in Sec.~\ref{sec:ground-state-prep}. We also briefly explain the noise model used in our simulations.

\subsection{Time evolution of a SU(2) unit cell}
For a unit cell of SU(2), the Hamiltonian can be obtained from Eq.~(\ref{eq:qubitkinetic})--(\ref{eq:qubitelectric}) using $N=2$. For time evolution, $\mu$ is taken to be zero. The Hamiltonian can then be split into two mutually non-commuting terms:
\begin{align}
     \hat{H}_{diag}=&\left(2m+\frac{3g^2}{8}\right)+ \frac{m}{2}(\hat{\sigma}_{3}^{z}+\hat{\sigma}_{4}^{z}
   -\hat{\sigma}_{1}^{z}-\hat{\sigma}_{2}^{z})-\frac{3g^2}{8}\hat{\sigma}_{1}^{z}\hat{\sigma}_{2}^{z}\,, \label{eq:h-diag}
\end{align}
\begin{equation}
    \hat{H}_{non-diag}=-\frac{1}{4}(\hat{\sigma}_{1}^{x}\hat{\sigma}_{2}^{z}\hat{\sigma}_{3}^{x}+\hat{\sigma}_{1}^{y}\hat{\sigma}_{2}^{z}\hat{\sigma}_{3}^{y}+\hat{\sigma}_{2}^{x}\hat{\sigma}_{3}^{z}\hat{\sigma}_{4}^{x}+\hat{\sigma}_{2}^{y}\hat{\sigma}_{3}^{z}\hat{\sigma}_{4}^{y})\,, \label{eq:h-nondiag}
\end{equation}
with $\hat{H}_{unit-cell} = \hat{H}_{diag} + \hat{H}_{non-diag}$. Note that all terms in $\hat{H}_{diag}$ are diagonal and can be implemented using single-qubit $z-$rotation $R_Z(\varphi) = e^{-i\varphi\hat{\sigma}^z}$ and two-qubit entangling gate $R_{ZZ}(\varphi) = e^{-i\varphi\hat{\sigma}^z\otimes\hat{\sigma}^z }$. On the other hand, all terms in $\hat{H}_{non-diag}$ are three-body non-diagonal terms, which leads to three-body rotation gates that need to be decomposed into single-qubit parametrized rotations and CNOT gates.

\begin{figure}[ht]
    \centering
    \includegraphics[width=1.0\linewidth]{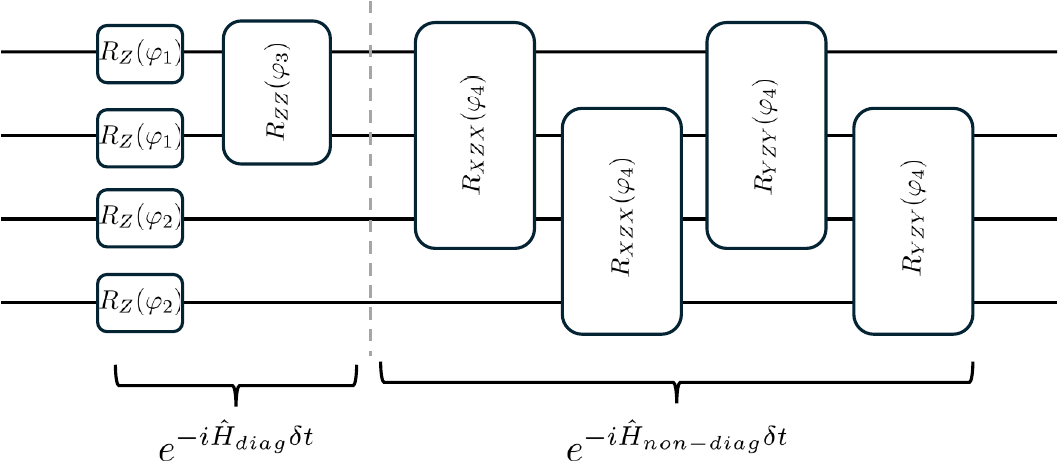}
    \caption{Circuit for a single trotter step for SU(2) unit cell in (1+1)-D. The different angles $\varphi_i$ are the rotation angles that depend on the coefficient of the terms that appear in the Hamiltonian and the trotter time step $\delta t$. In this particular case: $\varphi_1 = -\varphi_2 = m\delta t/2, \varphi_3=-3g^2\delta t/8, \varphi_4 = -\delta t/4$. Rotation gates are defined as $R_P(\varphi) = e^{-iP\varphi}$, where $P$ is a Pauli string, e.g,. $P=X\otimes Z\otimes X$, which by abuse of notation we write as $P = XZX$.}
    \label{fig:trotter-circuit}
\end{figure}

\begin{figure}[ht]
    \centering
    \includegraphics[width=1.0\linewidth]{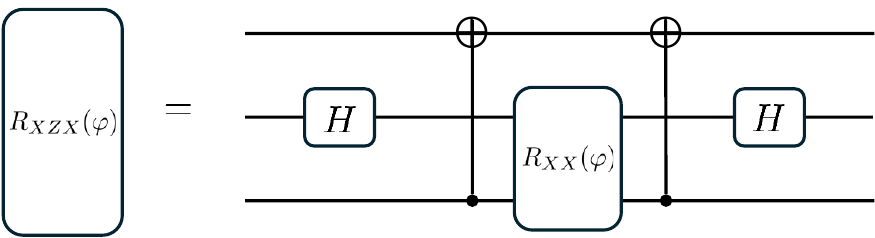}
    \caption{Decomposition of a three body rotation gate originating from a term in the Hamiltonian $\hat{H}_{non-diag}$ is shown here. The decomposition contains three two-body gates.}
    \label{fig:three-body-decomp}
\end{figure}

For the unit cell, we can now define one trotter step as 
\begin{equation}
    e^{-i\hat{H}\delta t} \approx e^{-i\hat{H}_{diag}\delta t}\,\cdot\,e^{-i\hat{H}_{non-diag}\delta t} \label{eq:trotter-single-decomp}
\end{equation}
where each term can be decomposed using the specific expressions given in Eq.~(\ref{eq:h-diag}) and (\ref{eq:h-nondiag}). The circuit to implement the right hand side of Eq.~(\ref{eq:trotter-single-decomp}) is shown in Fig.~\ref{fig:trotter-circuit}. Each three-body term in Eq.~(\ref{eq:h-nondiag}) requires three entangling gates (as shown in Fig.~\ref{fig:three-body-decomp}), which leads to a total of 12 two-body entangling gates. One additional two-body gate comes from the last term of Eq.~(\ref{eq:h-diag}) in implementing $e^{-i\hat{H}_{diag}\delta t}$. So, in total there are 13 two-body gates in a single trotter step, without performing any circuit reduction technique. 

As a noise model, a two-qubit depolarizing noise channel is added after each application of a two-qubit entangling gate, which results in a mixed state. 

\subsection{Ground state preparation}
\begin{figure}[ht]
    \centering
    \includegraphics[width=1.0\linewidth]{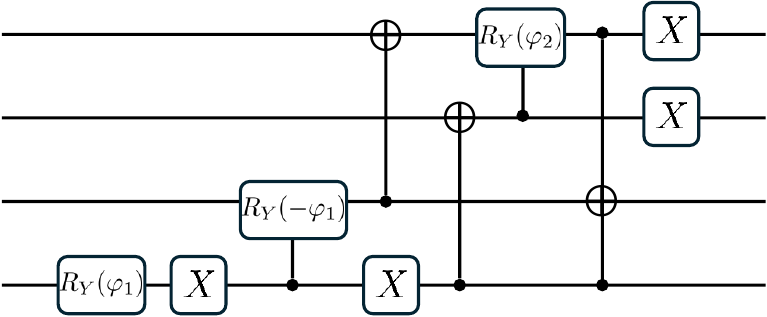}
    \caption{Circuit to prepare ground state of a SU(2) unit cell Hamiltonian. $X$ is the Pauli $\hat{\sigma}^x$ operator and $R_Y(\varphi)$ is a single qubit $y$ rotation.}
    \label{fig:meson-mass-circuit}
\end{figure}

The ground state preparation circuit is taken from Ref.~\cite{atas2021} and is shown in Fig.~\ref{fig:meson-mass-circuit}. This circuit prepares the ground state for the SU(2) unit cell Hamiltonian with parameters $m=0.5, g^2=0.5$.

The variational optimization is done in the presence and absence of noise using a gradient-free Bayesian direct search algorithm \cite{singh2024pybads}.

\subsection{Noise model}\label{appsec:noise-model}
In the circuits illustrated in this section, noise is applied to only two-qubit entangling gates. In both time-evolution and ground state preparation circuits, all three-qubit gates are decomposed in terms of single- and two-qubit gates. Each two-qubit gate is followed by a two-qubit local depolarizing noise channel with strength $\lambda_d$. For a two-qubit density matrix $\hat{\rho}_{2}$, the depolarizing noise channel $\mathcal{E}_{\lambda_d}$ is given by
\begin{equation}
    \mathcal{E}_{\lambda_d}(\hat{\rho}_2) = (1-\lambda_d)\hat{\rho}_2 + \frac{\lambda_d}{4}\hat{I}_{4\times 4}\;,
\end{equation}
where $\hat{I}_{4\times 4}$ is the $4\times4$ identity matrix. For a generic density matrix $\hat{\rho}$ (with more than two qubits), a two-qubit depolarizing channel applied on qubits $i$ and $j$ yields
\begin{equation}
   \mathcal{E}_{\lambda_d}^{i,j}(\hat{\rho}) = \left(1- \frac{15\lambda_d}{16} \right) \hat{\rho} \,+ \frac{\lambda_d}{16} \sum_{\substack{\hat{P}_i,\hat{P}_j \in \{\hat{\sigma}_x,\hat{\sigma}_y,\hat{\sigma}_z,\hat{I}\} \\ \hat{P}_i \hat{P}_j\neq \hat{I}\hat{I}}} \hat{P}_i\hat{P}_j \hat{\rho} \hat{P_j}\hat{P_i}\;.
\end{equation}
In the simulations with varying noise levels, the depolarizing noise channel strength $\lambda_d$ is changed. A larger value of $\lambda_d$ denotes a stronger noise channel.




\clearpage

\bibliography{biblio}

\end{document}